\documentclass[longauth]{aa}   

\usepackage{natbib}               
\usepackage{graphicx}
\usepackage{txfonts}
\usepackage{media9}
\usepackage[normalem]{ulem}

\usepackage[usenames,dvipsnames]{color} 
\usepackage{todonotes}
\usepackage{hyperref}
\hypersetup{%
  colorlinks=true,
  linkcolor=blue,
  urlcolor=blue,
  citecolor=blue,
  }

\makeatletter
\renewcommand*\aa@pageof{, page \thepage{} of \pageref*{LastPage}}
\makeatother

\newcommand*\kms{\ensuremath{\textrm{km}\,\textrm{s}^{-1}}}
\newcommand{\am}{\ensuremath{^{\prime}}}
\newcommand{\as}{\ensuremath{^{\prime\prime}}}
\newcommand{\hinode}{Hinode}
\newcommand{\solo}{Solar Orbiter}

\begin{document}

   \title{A multiple spacecraft detection of the 2 April 2022 M-class flare and filament eruption during the first close \solo\ perihelion}


\titlerunning{Multiple spacecraft detection of the 2 April 2022 M-class flare and filament eruption} \authorrunning{Janvier et al.}

   \author{
  	 M. Janvier  
  	 \inst{\ref{i:estec},\ref{i:ias},\ref{i:cogi}}
  	 \and
  	 S. Mzerguat
  	 \inst{\ref{i:ias}}
          \and
          P.~R.~Young 
          \inst{\ref{i:gsfc},\ref{i:numbria}}
          \and
          É. Buchlin
          \inst{\ref{i:ias}}
          \and
          A. Manou
          \inst{\ref{i:ias}}
          \and
          G. Pelouze
          \inst{\ref{i:ias}}
          \and
          D.~M.~Long 
          \inst{\ref{i:mssl},\ref{i:qub}}
          \and
          L. Green
          \inst{\ref{i:mssl}}
          \and
          A. Warmuth
          \inst{\ref{i:aip}}
          \and
          F. Schuller
          \inst{\ref{i:aip}}
          \and
          P. Démoulin
          \inst{\ref{i:lesia},\ref{i:cogi}}
          \and 
           D.~Calchetti
           \inst{\ref{i:mps}} 
           \and         
           F.~Kahil
          \inst{\ref{i:mps}} 
          \and 
          L.~Bellot Rubio
          \inst{\ref{i:iaa}}
          \and
          S. Parenti
          \inst{\ref{i:ias}} 
          \and
          S. Baccar
          \inst{\ref{i:ias}}
          \and
          K. Barczynski
          \inst{\ref{i:pmod},\ref{i:ETH}}
          \and
          L.~K.~Harra
          \inst{\ref{i:pmod},\ref{i:ETH}}
          \and
          L. A. Hayes
          \inst{\ref{i:estec}} 
          \and
          W. T. Thompson
          \inst{\ref{i:adnet}}
          \and
        	 D. Müller
         \inst{\ref{i:estec}}             
        	\and
          D. Baker 
          \inst{\ref{i:mssl}}
          \and
          S. Yardley 
          \inst{\ref{i:reading}, \ref{i:mssl}, \ref{i:dipc}} 
          \and
        D. Berghmans
          \inst{\ref{i:rob}}             
          \and 
          C. Verbeeck
          \inst{\ref{i:rob}}
        \and
        P.J. Smith
          \inst{\ref{i:mssl}}  
          \and
          H. Peter 
          \inst{\ref{i:mps}}  
        \and
        R. Aznar Cuadrado   
          \inst{\ref{i:mps}}
          \and
          S. Musset 
          \inst{\ref{i:estec}}
          \and 
          D. H. Brooks
          \inst{\ref{i:gmu}}
        \and
        L. Rodr\'{\i}guez
          \inst{\ref{i:rob}}
        \and
            F. Auchère
            \inst{\ref{i:ias}}
          \and
            M. Carlsson
            \inst{\ref{i:oslo}}
          \and
            A. Fludra
            \inst{\ref{i:ral}}
          \and
            D. Hassler
            \inst{\ref{i:swri}}
          \and
            D. Williams
            \inst{\ref{i:esac}}
          \and
            M. Caldwell
            \inst{\ref{i:ral}}
          \and
            T. Fredvik
            \inst{\ref{i:oslo}}
          \and
            A. Giunta
            \inst{\ref{i:ral}}
          \and
            T. Grundy
            \inst{\ref{i:ral}}
          \and
            S. Guest
            \inst{\ref{i:ral}}
          \and
            E. Kraaikamp
            \inst{\ref{i:rob}}
          \and
            S. Leeks
            \inst{\ref{i:ral}}
          \and
            J. Plowman
            \inst{\ref{i:swri}}
          \and
            W. Schmutz
            \inst{\ref{i:pmod}}
          \and
            U. Schühle
            \inst{\ref{i:mps}}
          \and
            S.D. Sidher 
            \inst{\ref{i:ral}}
          \and
            L. Teriaca
            \inst{\ref{i:mps}}
        \and  
        S.K.~Solanki
            \inst{\ref{i:mps}} 
        \and
        J.C.~del Toro Iniesta
            \inst{\ref{i:iaa}}     
        \and       
        J.~Woch
            \inst{\ref{i:mps}} 
            \and 
        A.~Gandorfer
            \inst{\ref{i:mps}} 
            \and
        J.~Hirzberger
            \inst{\ref{i:mps}} 
            \and
        D.~Orozco Su\' arez
            \inst{\ref{i:iaa}} 
            \and
        T.~Appourchaux
            \inst{\ref{i:ias}} 
            \and 
        G.~Valori
            \inst{\ref{i:mps}} 
            \and
        J.~Sinjan
            \inst{\ref{i:mps}} 
            \and               
        K.~Albert
            \inst{\ref{i:mps}} 
            \and
        R.~Volkmer
            \inst{\ref{i:kis}} 
          }

   \institute{
   European Space Agency, ESTEC, Noordwijk, The Netherlands \label{i:estec} \\
              \email{miho.janvier@esa.int}
    \and
    Université Paris-Saclay, CNRS,  Institut d'Astrophysique Spatiale, 91405, Orsay, France\label{i:ias}
    \and
       NASA Goddard Space Flight Center, Greenbelt, MD 20771, USA\label{i:gsfc}
    \and
       Northumbria University, Newcastle upon Tyne, NE1 8ST, UK\label{i:numbria}
    \and
        Mullard Space Science Laboratory, UCL, Holmbury St. Mary, Dorking, Surrey, RH5 6NT, UK\label{i:mssl}
    \and
        Leibniz-Institut f\"ur Astrophysik Potsdam (AIP), An der Sternwarte 16, 14482 Potsdam, Germany\label{i:aip}
    \and
        LESIA, Observatoire de Paris, Universit\'e PSL, CNRS, Sorbonne Universit\'e, Univ. Paris Diderot, Sorbonne Paris Cit\'e, 5 place Jules Janssen, 92195 Meudon, France\label{i:lesia}
     \and
        Laboratoire Cogitamus, rue Descartes, 75005 Paris, France \label{i:cogi}
     \and
        Max Planck Institute for Solar System Research, Justus-von-Liebig-Weg 3, 37077 Göttingen, Germany
        \label{i:mps}
    \and
        Instituto de Astrof\'isica de Andaluc\'ia (IAA-CSIC), Apartado de Correos 3004, E-18080 Granada, Spain\label{i:iaa}
    \and
        Physikalisch Meteorologisches Observatorium Davos, World Radiation Center, 7260 Davos, Switzerland\label{i:pmod}
        \and
         ETH Z\"urich, Institute for Particle Physics and Astrophysics , Wolfgang-Pauli-Strasse 27, 8093 Z\"urich, Switzerland \label{i:ETH}
    \and
        Astrophysics Research Centre, School of Mathematics and Physics, Queen’s University Belfast, University Road, Belfast, BT7 1NN, Northern Ireland, UK\label{i:qub}
    \and
        Department of Meteorology, University of Reading, Reading, UK\label{i:reading}
     \and
        Donostia International Physics Center (DIPC), Paseo Manuel de Lardizabal 4, 20018 San Sebastián, Spain\label{i:dipc}
     \and 
        College of Science, George Mason University, Fairfax, VA 22030, USA \label{i:gmu}
    \and
        RAL Space, UKRI STFC Rutherford Appleton Laboratory, Harwell, Didcot, OX11 0QX, UK
        \label{i:ral}
    \and   
        Solar-Terrestrial Centre of Excellence – SIDC, Royal Observatory of Belgium, Ringlaan -3- Av. Circulaire, 1180 Brussels, Belgium
        \label{i:rob}
    \and
        Leibniz-Institut für Sonnenphysik, Sch\" oneckstr. 6, D-79104 Freiburg, Germany
        \label{i:kis}  
    \and
        Adnet Systems, Inc., NASA Goddard Space Flight Center, Code 671, Greenbelt, MD 20771, USA
        \label{i:adnet}
       \and
       Institute of Theoretical Astrophysics, University of Oslo, Oslo, Norway
         \label{i:oslo}
         \and
         Southwest Research Institute, Boulder, CO, USA
         \label{i:swri}
         \and
            European Space Agency, ESAC, Villanueva de la Cañada, Spain \label{i:esac}  }
   \date{Received ...; accepted ...}

  \abstract
   {The \solo\ mission completed its first remote-sensing observation windows in the spring of 2022. On 2 April 2022, an M-class flare followed by a filament eruption was seen both by the instruments on board the mission and from several observatories in Earth's orbit, providing an unprecedented view of a flaring region with a large range of observations.}
   {We aim to understand the nature of the flaring and filament eruption events via the analysis of the available dataset. The complexity of the observed features is compared with the predictions given by the  standard flare model in 3D.}
   {In this paper, we use the observations from a multi-view dataset, which includes extreme ultraviolet (EUV) imaging to  spectroscopy and magnetic field measurements. These data come from the {Interface Region Imaging Spectrograph}, the {Solar Dynamics Observatory}, \hinode, as well as several instruments on \solo. }
   {The large temporal coverage of the region allows us to analyse the whole sequence of the filament eruption starting with its pre-eruptive state. Information given by spectropolarimetry from SDO/HMI and \solo\ PHI/HRT shows that a parasitic polarity emerging underneath the filament is responsible for bringing the flux rope to an unstable state. As the flux rope erupts, \hinode\ EIS captures blue-shifted emission in the transition region and coronal lines in the northern leg of the flux rope prior to the flare peak. This may be revealing the unwinding of one of the flux rope legs. At the same time, \solo\ SPICE captures the whole region, complementing the Doppler diagnostics of the filament eruption. Analyses of the formation and evolution of a complex set of flare ribbons and loops, of the hard and soft X-ray emissions with STIX, show that the parasitic emerging bipole plays an important role in the evolution of the flaring region.
}
   {The extensive dataset covering this M-class flare event demonstrates how important multiple viewpoints and varied observations are in order to understand the complexity of flaring regions. While the analysed data are overall consistent with the standard flare model, the present particular magnetic configuration shows that surrounding magnetic activity such as nearby emergence needs to be taken into account to fully understand the processes at work. This filament eruption is the first to be covered from different angles by spectroscopic instruments, and provides an unprecedented diagnostic of the multi-thermal structures present before and during the flare. This complete dataset of an eruptive event showcases the capabilities of coordinated observations with the \solo\ mission.}


   \maketitle
%


\section{Introduction}
\label{sec-Introduction}

\begin{figure*}
\centering
\includegraphics[width=\textwidth]{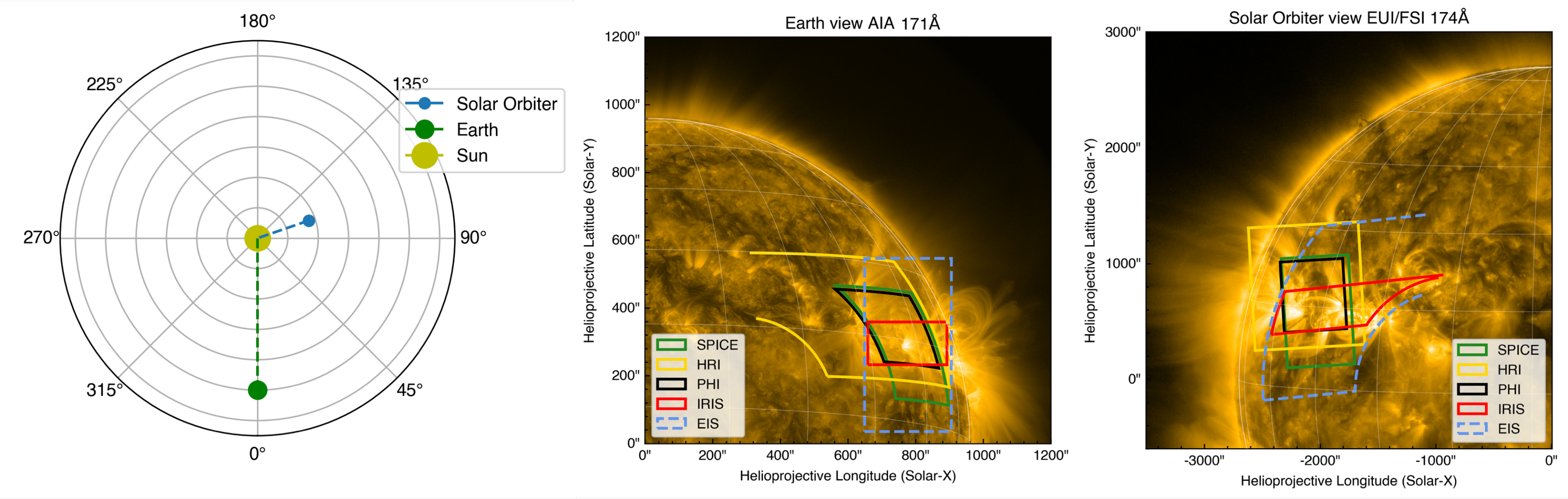}
\caption{Overview of the Solar Orbiter position and instruments fields of view. (Left) Positions of \solo\ (blue) and Earth (green) on 2 April 2022. The fields of view of the different instruments observing the flare, namely: IRIS and \hinode/EIS, \solo\ EUI/HRI, PHI/HRT and SPICE, as viewed from Earth (middle) overlaid on top of an AIA 171 \AA\ image, and as viewed from \solo\ (right) overlaid on top of an EUI FSI 174 \AA\ image.}

     \label{solo-position}
\end{figure*}

Solar flares are amongst the most energetic events taking place in our Solar System. They correspond to the release of free magnetic energy stored in the Sun's atmosphere. This process leads to a variety of phenomena following the energy deposition in various forms \citep[see][for a review of flare observations]{Benz2017}. These phenomena are seen in the different layers of the solar atmosphere, from Extreme UV (EUV) emitting loops to chromospheric, and sometimes photospheric, flare ribbons. EUV imaging also shows the formation of hot coronal loops as well as, in some cases, indications of the ejection of a twisted magnetic field structure called a flux rope \citep[e.g.][]{Cheng2014, Kumar2014}. Both loops and ribbons indicate a change in the magnetic field configuration, as shown for example by the location of electric current ribbons (derived directly from photospheric magnetograms). These currents are spatially located where the EUV ribbons are seen \citep{Janvier2014}.
Non-thermal hard X-ray sources are routinely seen at the footpoints of flare loops \cite[e.g.][]{Krucker2011}, and in some cases at the top of flare loops as well \citep[e.g.][]{Masuda1994}. Generally, the accelerated electrons that are associated with these sources are considered as a major mechanism to heat the chromospheric and coronal plasma \cite[e.g.][]{Warmuth2020a}.

The ultraviolet (UV) range gives access to emission lines that formed at all layers of the atmosphere from the chromosphere at 10~000\,K to the flaring corona at 20\,MK. UV slit spectrometers such as the Interface Region Imaging Spectrograph \citep[IRIS,][]{DePontieu2014}, the EUV Imaging Spectrometer (EIS) on \hinode\ \citep{Culhane2007}, and the Spectral Imaging of the Coronal Environment (SPICE) on \solo\ \citep{SPICE2020} yield line profiles for studying plasma dynamics as well as temperature, density, and element abundance diagnostics. Specific topics studied with spectrometer data include chromospheric evaporation at flare ribbon sites \citep{2009ApJ...699..968M}, enhanced line broadening during the flare build-up \citep{2008ApJ...679L.155I, Harra2013}, and large Doppler shifts during filament eruptions \citep{2023AdSpR..71.1900Y}. A review of spectrometer flare results is given by \citet{Milligan2015}. A disadvantage of slit spectrometers is that spatial scans of active regions can take much longer than the evolutionary timescales of eruptive flares. By leaving the slit in a fixed position, high cadence observations can be obtained but only for a single part of the active region. An example is presented here where the EIS slit was ideally positioned to capture the filament eruption. Furthermore, diagnostics from spectrometers are strongly constrained by the projection effect of the sources. While some evolution of parameters such as sustained Doppler shifts can be decorrelated from the projection by taking into account the solar rotation \citep[e.g.][]{Demoulin2013, Baker2017}, most observations would benefit from better constrained measurements obtained with two different lines of sight, which is the case for the present event.

With all the different observational constraints, a flare model emerged over the years, starting from the CSHKP model \citep{Carmichael1964, Sturrock1966, Hirayama1974, Kopp1976}. Different views of this model have been proposed since, to include the various features seen in the observations. However, a key piece missing from most of these models is the 3D aspect, which is essential to explain the morphology of flare ribbons and understand the 3D nature of the magnetic field reconnection leading to the formation and apparent dynamics of flare loops \citep[e.g.][]{Aulanier2010, Dudik2014}. Using numerical simulations, several authors have analysed the consequences of 3D magnetic reconnection during flares in an effort to complete our understanding of the observations \citep[e.g.][]{Aulanier2012, Kliem2013, Amari2018}. The flare reconnection produces a feedback effect since it increases the poloidal flux of the rope, aiding its acceleration away from the Sun. This can in turn affect the flare reconnection rate \citep{Lin2000}. If a flux rope is pre-existing, the specific configuration of the rope can influence how its eruption proceeds \citep[more discussions can be found in][]{Patsourakos2020}. Therefore, a strong interest lies in the formation and characterisation of pre-flare flux ropes.

The formation and initiation of a coronal mass ejection (CME) is also the subject of various scenarios, starting with the existence or not of a flux rope prior to the eruption, and different trigger processes such as the torus instability (discussed below) or the breakout model \citep{Antiochos1999}. The physical processes involved in the onset of an eruption involve both ideal and non-ideal processes acting on non-potential fields. A theoretical view is a
pre-eruptive magnetic field that is formed by a sheared arcade where reconnection is initiated and drives the eruption \citep{Karpen2012}. In other models, the pre-eruptive configuration is a flux rope. When a flux rope is present, the eruption can be driven by a loss of equilibrium \citep{Vantend1978}, a flux rope catastrophe \citep{Forbes1991}, or the torus instability \citep{Kliem2006}, which are different views of the same physical phenomena \citep{Demoulin2010}.

Observationally, flux cancellation due to photospheric motions and flux emergence are among the most recognised processes at the origin of the filament shearing and destabilisation. They result in local reconnection events that partially dissipate the accumulated magnetic energy and partly change the magnetic topology. In the corona, this is manifested by EUV brightenings and fading, both in the main body and at the footpoints of the filament. During the pre-eruption phase, the filament body can show deformation and helical motion, as well as oscillations \citep[see][for a general review of these processes]{Parenti2014}. Furthermore, non-thermal velocities in the corona have long been measured in the early phases of an eruption. The non-thermal velocity is inferred from a line-width enhancement above the thermal width in an emission line. These enhancements can be seen minutes before a flare begins and in some cases hours before, and can be related to pre-flare activity turbulence or multi-directional flows. For example, \cite{Harra2013} looked at three large eruptive flares and found that the enhanced non-thermal velocity in the pre-flare stages was found to lie away from what would become the flare footpoints, and on what would become the footpoints of the dimming regions. This provides further evidence that pre-flare non-thermal enhancements in the corona can be related to a magnetic flux rope slowly reaching a threshold leading to a subsequent eruption.

For flux ropes that formed low in the solar atmosphere, likely through the process of flux cancellation \citep{Vanballegooijen1989,Green2011}, the specific configuration can contain field lines that graze the photosphere and have a concave-up shape, forming so-called bald patches \citep[BPs:][]{Titov1993, Muller2008}. In BP ropes, the underside of the rope reaches down into the denser plasma layers (photosphere, chromosphere) where helical field lines have concave-up sections and are therefore anchored by the line tying condition. 

There are observational and modelling studies which show that although flux ropes may form with a BP configuration via the process of flux cancellation, magnetic reconnection can then occur within the BP separatrix layer or at the separator present between two BPs \citep{Titov1999}. The result of this is that the flux rope transitions from a BP configuration to one in which the underside of the rope is higher in altitude with quasi-separatrix layers (QSLs) separating it from the neighbouring arcade \citep{Demoulin1996b}. 
The thinnest region of the QSLs is known as a hyperbolic flux tube \citep[HFT;][]{Titov2002}. A thin current layer is naturally formed at the QSLs, and especially at the HFT, during the flux rope evolution and an even stronger one is built when the flux rope gets unstable \citep{Aulanier2010,Green2011,Savcheva2012,Zuccarello2015}. Then, magnetic reconnection occurs, aiding the eruption of the rope.

Flares can also inject particles and magnetic flux into the heliosphere. Solar energetic particles (SEPs) originating from flaring regions can be injected via different mechanisms, from the flare site to the shocks that can accompany the release of a CME. As flares are space weather inducing events \citep{Temmer2021}, a strong interest remains in understanding the mechanisms underlying the events leading up to flares.

One of the objectives of the \solo\ mission \citep{Muller2020, Garcia2021}, launched in February 2020, is to understand how the solar variability on different scales affects the heliosphere. Since the launch, the probe has recorded a variety of eruptive solar flares \citep[
][this issue]{Berghmans2023}. In this paper, we present the first comprehensive study of an eruptive solar flare seen by the remote-sensing instruments on board the mission, completed by near Earth assets. 

The paper is outlined as follows. In Sect. \ref{sec-Overview}, we present an overview of the set of observations from the 2 April 2022 eruptive flare event. In Sect. \ref{sec-Precursor}, we discuss the precursor phase of the flare taking place a few days to a few hours before the eruption. 
{The flare loops and ribbons are analysed with the spectroscopic observations taken by 
\solo/SPICE (Sect. \ref{sec-SPICE}), the filament eruption by \solo/SPICE and \hinode/EIS (Sect. \ref{sec-Filament}),  while the hot plasma and energetic electrons are studied by \solo/STIX (Sect. \ref{sec-Sources}). In Sect. \ref{sec-Discussions} all observations are placed in the context of our present understanding of 3D MHD flare/eruption models. Finally, our} conclusions are given in Sect.~\ref{sec-Conclusions}.

  \begin{figure}
  \centering
  \includegraphics[width=0.5\textwidth]{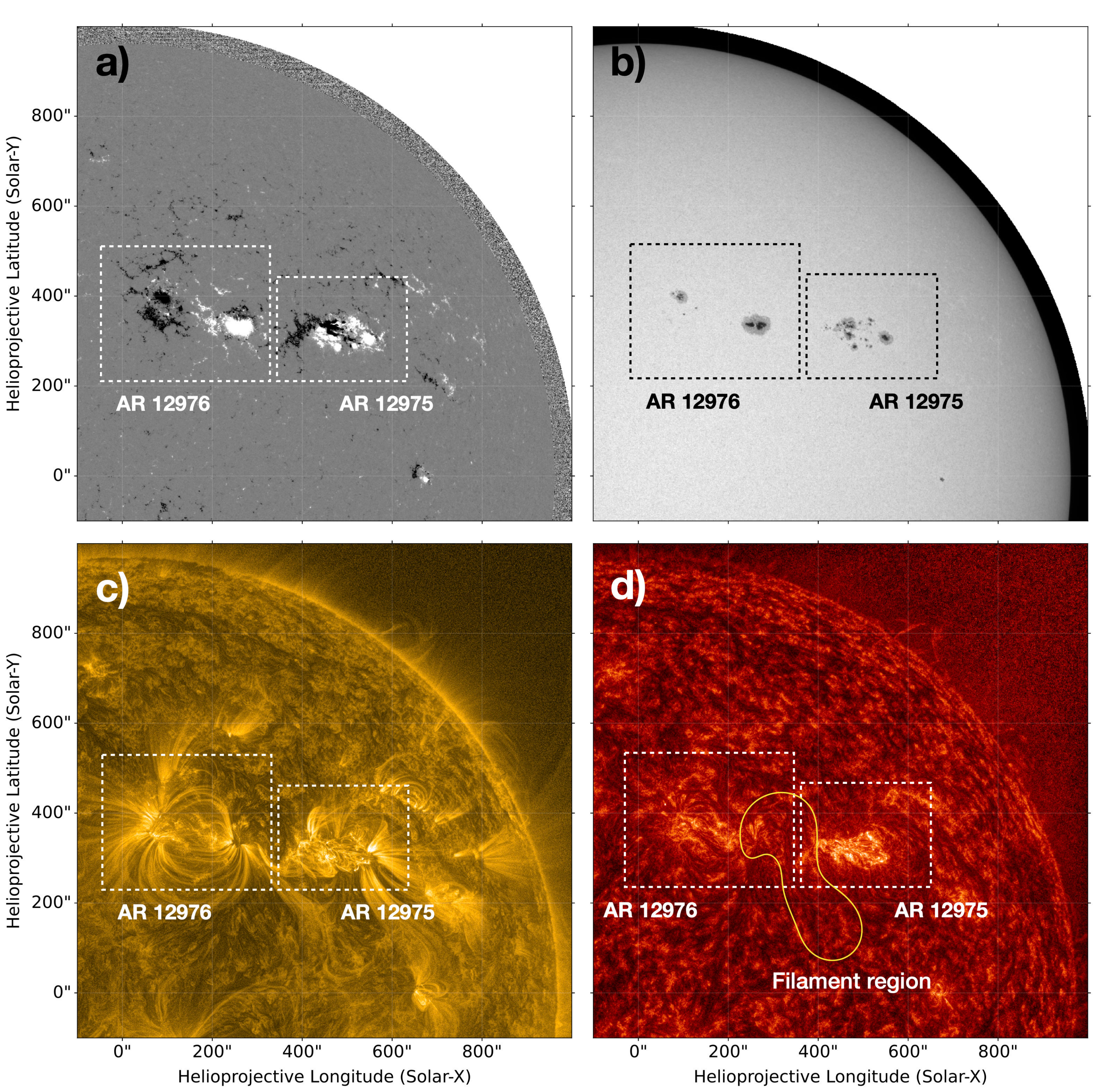}
      \caption{Overview of the two active regions AR 12975 and AR 12976 seen from the Sun-Earth line on 30 March 2022 during the pointing decision meeting. The different views correspond to the (a) HMI magnetogram and (b) continuum data, (c) AIA 171\,\AA\ and (d) AIA 304\,\AA. A filament is present in between the two regions in the AIA 304\,\AA\ image (d) (yellow contour). The AIA images have been processed using the multi-scale Gaussian normalisation \citep{Morgan2014}.}
         \label{overview-activeregions}
  \end{figure}

  \begin{figure}
  \centering
  \includegraphics[width=0.5\textwidth]{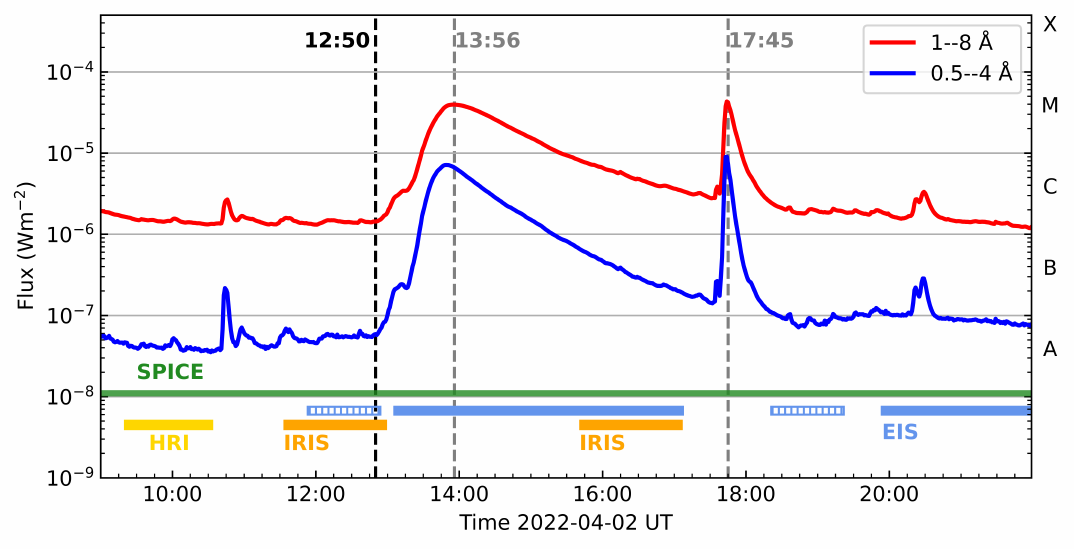}
      \caption{X-ray flux seen in the two GOES channels (1-8~\AA\ and 0.5-4~\AA) on 2 April 2022. The first flare corresponds to the filament eruption, which is the focus of this paper, while the second confined flare occurred later on in the same region. The vertical dashed black line shows the beginning of the impulsive phase of the first flare, while the two vertical dashed grey lines show the peaks of the two flares. Both flares were classified as M-class flares. The bottom of the graph shows the temporal coverage of the observations by the different optical instruments discussed in the paper, namely SPICE (green), EIS (blue), EUI/HRI (yellow) and IRIS (orange).
      }
         \label{xrayflux}
  \end{figure}

  \begin{figure*}
  \centering
  \includegraphics[width=\textwidth]{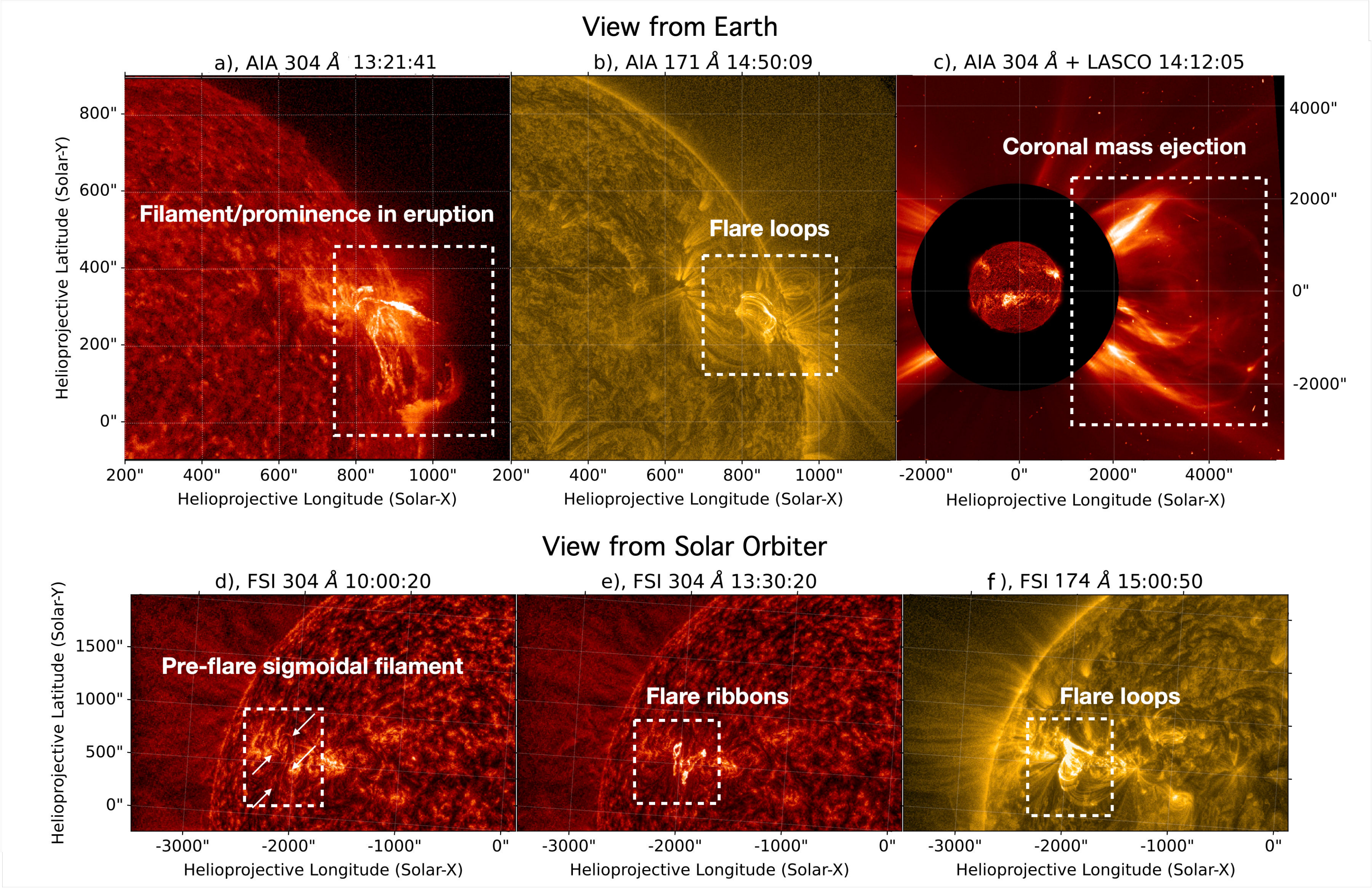}
      \caption{ Views of the flaring region from different imagers and at different times. The SDO/AIA 304\,\AA\ image from the impulsive phase of the event shows the prominence in eruption (a) while the SDO/AIA 171\,\AA\ image from the decay phase shows the flare loops (b). The filament eruption subsequently led to a CME detected by the SOHO/LASCO C2 coronagraph (c). Times are at Earth time. The \solo\ EUI-FSI 304\,\AA\ image shows the pre-flare filament and the flare ribbons (d and e), and the 174\,\AA\ image shows the flare loops (f). Times are at the observation time (321s before Earth time). The AIA images have been processed using the multi-scale Gaussian normalisation \citep{Morgan2014}.}
         \label{imageurs-euv}
  \end{figure*}

%
\begin{table*}
\caption{List of observations and associated individual events before and during the M3.9-class flare. All timings are given at Earth time.}             
\label{table-features}      
\centering          
\begin{tabular}{l l l l }     
\hline\hline       
                      
Approx. time (UT) & Observations & Event description & Notes\\ 
\hline                    
 From 1 April & PHI/HRT, HMI, Hinode/SP    & Flux emergence in between AR 12975 and 12976 & Fig. \ref{magnetograms} \\  

 From 1 April  & AIA    & Filament slow rise    & Figs. \ref{magnetograms},\ref{aia-looptop} \\
\hline
2 April & & & \\
  03:39 - 11:53 & EIS  &  Non-thermal velocities seen near & Fig.~\ref{vnt-eis}\\

  &   &  the filament northern root part  & \\

  09:24 - 10:39 & EUI/HRI, SPICE &  Overlying arcade above filament apparent motion    & Fig.~\ref{hri-spice-loops-stackplots}\\
  
  11:32 & IRIS  &  Northern part of the filament seen in IRIS/\ion{Mg}{ii}   & Fig.~\ref{iris}a\\
  
   12:50 & GOES, STIX  &  Start of the flare   & Figs.~\ref{xrayflux}, \ref{goes_stix_lc}\\

    13:10 & AIA  &  Filament fast rise   & Figs.~\ref{imageurs-euv}a, \ref{aia-looptop}\\

  13:00 -- 14:00 & EUI/FSI 304\,\AA, SPICE, STIX  & Flare ribbons, HXR sources, strong Doppler shifts    & Figs.~\ref{imageurs-euv}e,\ref{spice-intensities},\ref{spice-fexviiimaps},\ref{spice-doppler} \\

   13:05 -- 17:08 & EIS        & Sit and stare observations    & Figs.~\ref{fig.composite}, \ref{fig.movie1}, \ref{fig.he2-twist}, \ref{fig.multitemp}, \ref{fig.eis-aia}  \\

  13:56 & GOES, STIX  &  Peak of the flare   & Figs.~\ref{xrayflux}, \ref{goes_stix_lc}\\

  14:12 & SOHO/LASCO  &  Coronal mass ejection   & Fig.~\ref{imageurs-euv}c\\
  
    $\approx$ 14:50 & AIA, EUI/FSI 174\,\AA, SPICE &  Flare loops   & Figs.~\ref{imageurs-euv}b, f,\ref{spice-intensities},\ref{spice-fexviiimaps} \\
 
  15:39 & IRIS & Part of the northern filament leg still seen in \ion{Mg}{ii}   & Fig.~\ref{iris}b\\
  \hline                  
\end{tabular}
\end{table*}


\section{Overview of the observations}
\label{sec-Overview}
\subsection{Long-term Active Region SOOP}
\label{sec-Longterm}

In the first months of 2022, three remote-sensing windows of ten days each were targeting different science objectives\footnote{See \url{https://issues.cosmos.esa.int/solarorbiterwiki/display/SOSP/LTP06+Q1-2022} for more detailed information on each}. The third remote-sensing window during which the observations presented in this paper were conducted ran from 27 March 2022 until 6 April 2022. Then, \solo\ was situated between 0.323\,AU (first perihelion) and 0.391\,AU, as indicated in Fig.~\ref{solo-position}.  On 2 April 2022, the distance between {Solar Orbiter} and the Sun centre was 0.356\,AU. Because of this distance, the timings of the observations made at Solar Orbiter are different for the same events recorded at Earth. Due to the difference in light travel time to {Solar Orbiter} compared to the Earth, an observation time at the spacecraft needs be increased by 321\,s to be equivalent to the observing time for an Earth-orbiting spacecraft (\hinode, {Solar Dynamics Observatory} --- SDO). Throughout the paper, both times are indicated when relevant.

The \solo\ remote-sensing observations are designed to address the mission's science objectives, and executed by means of the Solar Orbiter observing plans \citep[SOOPs, see][]{Zouganelis2020}. During the first remote-sensing observation campaign of the nominal mission phase in March--April 2022, one of the SOOPs was dedicated to study the long term evolution of an active region 
(long-term AR follow-up\footnote{\url{https://issues.cosmos.esa.int/solarorbiterwiki/display/SOSP/R_SMALL_MRES_MCAD_AR-Long-Term}}) and was coordinated by L.\ Bellot Rubio. This corresponded to four full days of observations from 31 March 17:45\,UT to 4 April 16:20\,UT with a short-term pointing decision made a few days before the start of the observations. Coordinated observations were performed with \hinode\ and IRIS through the \href{https://www.isas.jaxa.jp/home/solar/hinode_op/hop.php?hop=0436}{\hinode\ Operation Plan (HOP) 436}.

Since the primary science objective for this SOOP was to look at the decay of an active region, and considering the state of the Sun a few days prior to the start of the observations, a choice was made to focus on the two nearby regions AR 12975 and AR 12976 (see Fig.~\ref{overview-activeregions} for different views of the region taken with the HMI and AIA instruments aboard SDO). While both active regions had started their decay, region 12976 showed a consistent leading polarity sunspot (see Fig.~\ref{overview-activeregions}(b) for the SDO/HMI continuum data), which became the focus of the pointing decision. Region 12975 was also in its decay phase but new flux emerged into the positive polarity between 27 March and 31 March. Due to the relatively large field of view chosen for the \solo\ EUI and SPICE instruments (see Fig.~\ref{solo-position}), the following polarity region of AR 12975 was also planned to be observed. 

\begin{figure*}
    \centering
    \includegraphics[width=\textwidth]{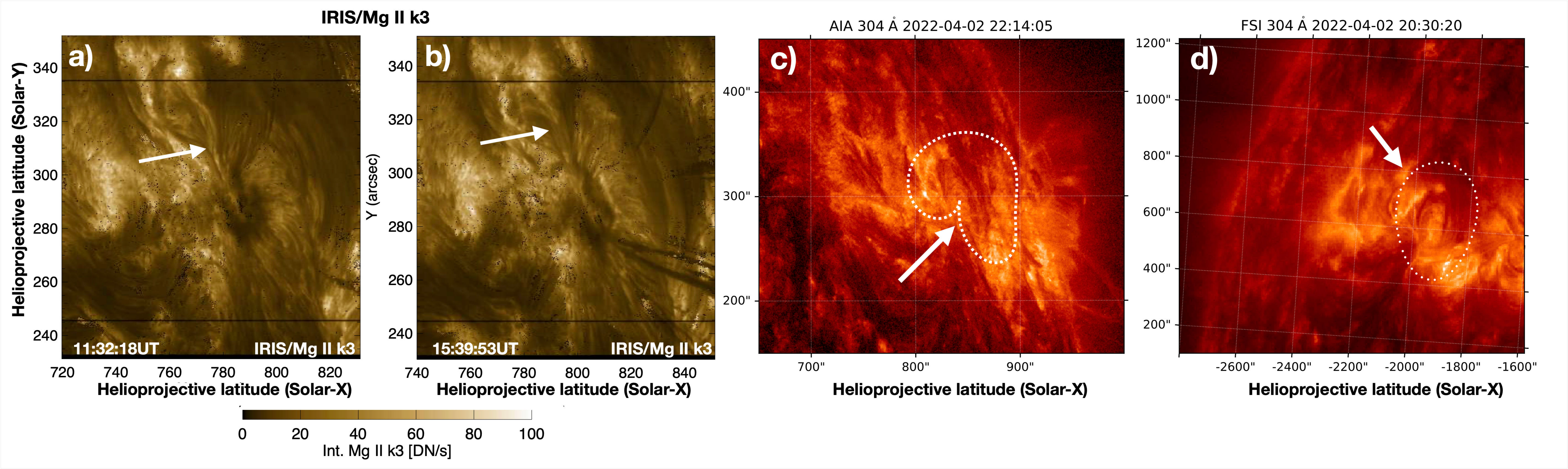}
    \caption{Views of the flaring region by IRIS with \ion{Mg}{II} k3 line before the flare peak (a), after flare peak (b), AIA 304\,\AA\ (c), and EUI/FSI 304\,\AA\ (d). The white arrows highlight the hook part of the filament, as well as the dotted contour lines in the AIA and EUI/FSI observations.}
    \label{iris}
\end{figure*}

\subsection{M-class flare on 2 April 2022}
\label{sec-M-class}

On 2 April 2022, two M-class flares were recorded between 12:00 and 22:00~UT from the region containing AR 12975 and 12976, as shown by the GOES X-ray Sensor (XRS) light curves in Fig.~\ref{xrayflux}. This figure also shows the availability of the {FUV/EUV} datasets presented in the paper at different times during the flares.
We focus on the first flare, which featured a filament eruption and CME  and had a GOES classification of M3.9. The filament was located between the two active regions, as indicated with the yellow contour shown in Fig.~\ref{overview-activeregions}~(d). The flare started around 12:50\,UT, and reached  a peak at 13:56\,UT in the low energy bands of GOES-16. The impulsive phase shows a two-step increase in the GOES/XRS light curve rather than a monotonic increase usually seen in flares.
The peak is followed by a long decrease, as often seen in eruptive flares with long-lasting flare loops \citep[e.g.][]{Harra2016}.

The flare was recorded by different EUV imaging instruments and a selection of images is shown in Fig.~\ref{imageurs-euv}. The SDO Atmospheric Imaging Assembly \citep[AIA,][]{Lemen2012}, recording the flare at 12\,s cadence in different passbands, clearly shows the filament/prominence in eruption (panel (a) in 304\,\AA) while flare structures such as flare loops are seen in the decaying phase of the flare (e.g. panel b in 171\,\AA). Hot flare emission was also recorded in soft X-rays by the \hinode\ X-Ray Telescope \citep[XRT,][]{Golub2007}. The subsequent CME was observed with different coronagraphs, such as the Large Angle Spectroscopic Coronagraph \citep[LASCO,][]{Brueckner1995}/C2 on board SoHO (panel c). Throughout the day on 2 April, the EUV Imager \citep[EUI,][]{Rochus2020} on board \solo\ took images \citep[see data release ][]{euidatarelease5} with the Full Sun Imager (FSI) in the 174\,\AA\ channel every 10\,min, and in the 304\,\AA\ channel every 30\,min. The EUI High Resolution Imagers (HRI$_{\mathrm{EUV}}$ and HRI$_{\mathrm{Lya}}$) recorded images of the region every 10\,s in the pre-flare phase between 09:19 and 10:34\,UT. We discuss in more detail the observations obtained from HRI$_{\mathrm{EUV}}$ in Sect. \ref{sec-Precursor}. For convenience, we simply write in the following EUI/HRI to refer to EUI/HRI$_{\mathrm{EUV}}$ as we do not discuss HRI$_{\mathrm{Lya}}$ further. 
The FSI 304\,\AA\ image shows the pre-existing filament and its sigmoidal shape (d) as well as the {flare ribbons (e) appearing after the flare onset.} The FSI 174~\AA\ image also shows the flare loops from the \solo\ point of view as shown in (f). These features are typical of eruptive flares and are expected in the framework of the 3D standard flare model \citep[see the review of][and references therein]{Janvier2017}.

\subsection{Polarimetric observations}
During the entire duration of the SOOP, the High Resolution Telescope (HRT) of the Polarimetric and Helioseismic Imager \citep[PHI,][]{2020A&A...642A..11S} on \solo\ tracked the leading spot of AR 12976 and part of the trailing polarity of AR1275 at a regular cadence of 30\,min. Each observation consisted of a full polarimetric scan through the \ion{Fe}{i} 6173 \AA\/ line at 5 wavelength positions plus a continuum point. The raw data were downlinked to Earth and processed on-ground with the standard PHI/HRT pipeline \citep{2022SPIE12189E..1JS}. This included the inversion of the data using the C-MILOS code \citep{2007A&A...462.1137O}, resulting in maps of the full vector magnetic field and line of sight velocity field at photospheric levels. The calibration of the data presented here is preliminary, as it does not yet include compensation for wavefront errors due to thermal loads on the heat rejection entrance window of the instrument (Kahil et al., this issue). This, however, does not influence the longitudinal magnetic field maps, in particular the polarity of the field, which is the main parameter of interest for the present study.  

Coordinated observations with the spectropolarimeter \citep[SP,][]{2013SoPh..283..579L} on the \hinode\ Solar Optical Telescope \citep{2007SoPh..243....3K} allowed to extend the coverage of the AR evolution by several days before the start of the SOOP. On 2 April 2022, the leading spot of AR 12976 was scanned by the \hinode/SP in the normal map mode at 03:40, 07:39, 11:55 and 18:24~UT, covering a FOV of $80\arcsec \times 80\arcsec$ with a pixel size of 0\farcs16. Additionally, a wider normal map covering both AR 12975 and 12976 was taken at 14:06~UT (not shown). 

The PHI/HRT and \hinode/SP observations provide views of the flaring region, and in particular of emerging flux during the active region evolution, from two very different vantage points. In both cases, the region was recorded near the limb on 2 April, but PHI/HRT observed it close to the eastern limb and \hinode/SP close to the western limb. This difference in viewing angles led to apparent changes in the polarities of some areas over the FOV which are just due to projection effects. Having the two observations makes it possible to disentangle projection effects unambiguously, and will ultimately allow to apply magnetic stereoscopy to retrieve the vector magnetic field without the azimuth ambiguity that affects single vantage point measurements \citep[][Valori et al., this issue]{2022SoPh..297...12V}.

The PHI/HRT observations permit a simpler interpretation of the observations taken by EUI and SPICE, given that the three instruments have similar fields of view and observe the same target from the same vantage point, which makes the co-alignment of the different fields of view much easier. The PHI/HRT and EUI observations were manually aligned using the PHI/HRT longitudinal magnetograms and the EUI/HRI Ly-$\alpha$ images, which show good correspondence between photospheric magnetic flux concentrations and chromospheric enhanced emission. Further adjustments were made visually to correct for the slight offset between EUI/FSI and EUI/HRI.

\subsection{Spectroscopic observations}
\label{sec-Spectroscopic}

\begin{table}
        \centering
        \caption{Analysed spectral lines with EIS, IRIS and SPICE. \label{table-Ions}}
        \begin{tabular}{ c c c c }
             \hline\hline
             Instrument & spectral line & $\lambda$ [\AA] & $\log(T [\textrm{K}])$  \\ [0.5ex] 
             \hline
            IRIS & \ion{Mg}{ii}  & 2795.5 & 4.5 \\[.5ex]
             \hline
             & H Lyman\,$\beta$ & 1025.72 & 4.0 \\  [.5ex]
             
             & \ion{C}{iii} & 977.03 & 4.5 \\  [.5ex]
             
             SPICE & \ion{N}{iv}  & 765.15 & 5.1 \\ [.5ex]
             
             (\solo ) & \ion{O}{vi} & 1031.93 & 5.5 \\  [.5ex]
             
             & \ion{Ne}{viii} & 770.42 & 5.8 \\[.5ex]
             
             & \ion{Mg}{ix}    & 706.02 & 6.0 \\ [.5ex]

             & \ion{Fe}{xviii}  & 974.84 & 6.9 \\[.5ex]
             \hline 
             EIS (\hinode) & \ion{He}{ii}           & 256.32 & 4.9 \\[.5ex]
             & \ion{Fe}{viii} & 194.66 & 5.7 \\[.5ex]
             & \ion{Fe}{xii}           & 195.12 & 6.2 \\[.5ex]
             & \ion{Fe}{xv}  & 284.16 & 6.4 \\[.5ex]
             & \ion{Fe}{xxiv} & 192.03 & 7.3 \\[.5ex]
            \hline
            \hline
        \end{tabular}
\end{table}

A compelling aspect of the present observation is that the flare was monitored by the three available solar spectrometers situated at different positions. These are  \hinode /EIS \citep{Culhane2007} and IRIS \citep{DePontieu2014} from the Sun-Earth line, and SPICE on Solar Orbiter \citep{SPICE2020, Fludra2021}.
As this SOOP was planned as a coordinated campaign with the other missions \hinode\ and IRIS, both regions were part of their respective observing programmes under the observing time request HOP\,436.

\subsubsection{Chromospheric observations with IRIS}
\label{sec-Chromospheric-IRIS}

IRIS contributed to HOP 436 on 2 April, with Tiago M. D. Pereira as planner, with seven large, dense 320-step rasters, each spaced by around 4 hours (see Fig. \ref{xrayflux} and Table \ref{table-features} for the timing of these observations relative to the flare evolution). The period of the M3.9 flare rise and filament eruption was not observed, but there were two rasters obtained during 11:32--13:00~UT (before the flare/filament eruption) and  
15:39--17:07\,UT (after the flare/filament eruption) and we focus on these here.
The rasters have a 112\as\ $\times$119\as\ field-of-view (FOV) and a step cadence of 16.5\,s.
The IRIS raster data provide information about the solar chromosphere (\ion{Mg}{II} emission), upper solar chromosphere (\ion{C}{II} emission), as well as the transition region (\ion{Si}{IV} emission).
Here we focus on the k line of \ion{Mg}{ii} at 2795.5\,\AA\ and, in particular, images formed from the core of the line, which is referred to as k3 \citep[e.g.][]{2013ApJ...778..143P}.

We analysed the level-2 data, downloaded from the IRIS
database\footnote{\url{https://iris.lmsal.com/search/}}. To find the extrema of the \ion{Mg}{ii} spectral line (intensities) for the whole raster, we used the SolarSoftWare (SSW) routine
 iris\_get\_mg\_features\_lev2  \citep{2013ApJ...778..143P}.
Finally, the routine computed the intensity raster map of \ion{Mg}{ii} k3.

The raster maps of the \ion{Mg}{II} k3 line core also show the filament seen in the AIA images before the flare peak (Fig.~\ref{iris}a). Interestingly, we also see that the lowest portion of the northern part of the filament is still present after the flare, as indicated by the white arrow in Fig.~\ref{iris}b. 
This structure is also seen in the 304\,\AA\ filters of EUI (Fig.~\ref{iris}c) and AIA (Fig.~\ref{iris}d).
These observations show that while the filament eventually erupted, its northern part separated with some of the filament remaining anchored at the Sun. Other examples of partial eruption of filaments have been reported, and \citet{Gibson2006} have shown with an MHD numerical simulation that this is compatible with the initial filament plasma being caught in a flux rope.

\subsubsection{Upper chromosphere/TR/low corona observations with \solo\ SPICE}
\label{spicedata}

On 2 April, SPICE (instrument planners M. Janvier and É. Buchlin) took `dynamics' studies throughout the day, from 03:15 to 23:15\,UT. These comprised 160-step rasters with the $4\as \times 11\am$ slit and 5\,s exposure time, giving a raster cadence of 14\,min 10\,s.
Projecting the SPICE raster size to an Earth observer gives an area of $228\as \times 324\as$. Eight spectral windows were selected for the study corresponding to the emission lines shown in Table~\ref{table-Ions}.
We note that the \ion{Fe}{xviii} line is only observed in active regions and becomes strong during flares.

The level-2 FITS files from the data release 3.0 \citep{spice-doi} are used in the following.
Level-2 files are calibrated to physical units,
and are corrected for flat-field, dark current, geometric distortion, etc. \citep{SPICE2020}.
Depending on the number of expected spectral lines, each spectral window is fitted via a Gaussian or multiple Gaussians with the  \textsf{curve\_fit} function provided by Python SciPy library \citep{scipy}.
In areas with low signal-to-noise ratio, pixels are first convolved with a smoothing kernel, where this is needed for the fitting.

\subsubsection{Coronal observations with \hinode\ EIS}

The EIS Chief Observer for 2 April was D.~H.~Brooks, and HOP\,436 was supported by EIS with four runs of the study \textsf{HPW021VEL260x512v2} beginning at 03:39, 07:39, 11:53 and 18:21\,UT (see Fig. \ref{xrayflux} and Table \ref{table-features} for the timing of these observations relative to the flare evolution). The study resulted in one raster of size 260\as\ $\times$ 512\as\ with a duration of 61~min, discussed in Sect. \ref{sec-Flux_emergence} and shown in Fig.~\ref{vnt-eis}. 

The level-0 data were obtained from the Hinode/EIS archive\footnote{\url{http://solarb.mssl.ucl.ac.uk/SolarB/SearchArchive.jsp}}. The data were calibrated with the \verb+eis_prep+ standard calibration routine from the Solar Software IDL library.
This routine removes the CCD dark current, cosmic ray pattern, and hot and dusty pixels from detector exposures and provide the radiometric calibration to physical units (erg cm$^{-2}$ s$^{-1}$ sr$^{-2}$ \AA$^{-1}$). The slit tilt correction is also made.
The coronal emission line \ion{Fe}{xii} at 195.12\AA\ was analysed.
To perform the fitting of the spectral lines, the \verb+eis_auto_fit+ routine\footnote{Solar Software; eis\_auto\_fit.pro} was used.  A single Gaussian fit was applied to each spectrum from the \ion{Fe}{xii} emission line. 

Following the large flares on 30 and 31 March, the \hinode\ planning team called a Major Flare Watch for AR 12975 beginning on 2 April around 11\,UT. This led to  the study \textsf{Flare\_SNS\_v2} being run for two periods, 13:05--17:07\,UT and 19:53--23:55\,UT, and the first of these caught the flare studied in the present work. The study uses a fixed slit position (``sit-and-stare'' mode) and 900 exposures were obtained for each observing period. Each slit position has a 10\,s exposure followed by a 5\,s delay time, and the cadence averages to 16.1\,s. The 2\as\ slit is used and the spatial coverage along the slit is 152\as\ in the Y-direction. A comprehensive study of these observations is given in Sect. \ref{sec-eisdoppler}, 
for which the software and data for generating the figures are available in the GitHub repository \textsf{pryoung/papers/2023\_orbiter\_flare} \citep{peter_young_2023_7900949}.


During the first \textsf{Flare\_SNS\_v2} observing period, the \hinode\ spacecraft was tracking solar rotation and so the EIS slit drifted westwards by around 10\as. The $x$ location at the beginning of the sequence was $+917$\as, and the centre of the slit was at $y=+279$\as.

\begin{figure}
\centering
\includegraphics[width= 0.49\textwidth]{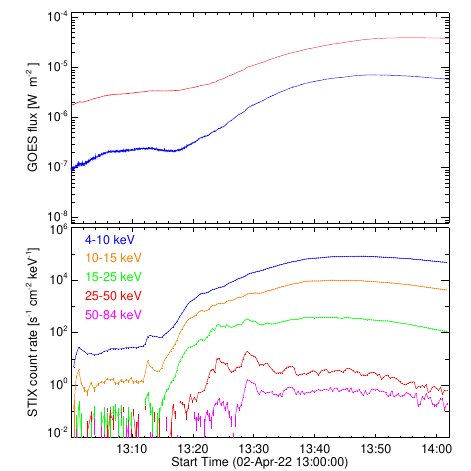}
\caption{X-ray lightcurves of the impulsive phase of the M3.9 flare. Top: GOES-16 soft X-ray fluxes. The GOES fluxes have been shifted in time to be consistent with the observations at \solo. Middle: STIX hard X-ray count rates in three broad energy bands that are dominated by thermal emission. Bottom: STIX HXR count rates in two energy bands that show non-thermal emission. The vertical dotted lines indicate the three major non-thermal peaks. A pre-event background has been subtracted from all STIX count rates. The high-energy count rates after the third peak are enhanced due to a detector saturation issue.} 
\label{goes_stix_lc}
\end{figure}

\begin{figure*}
    \centering
    \includegraphics[width=\textwidth]{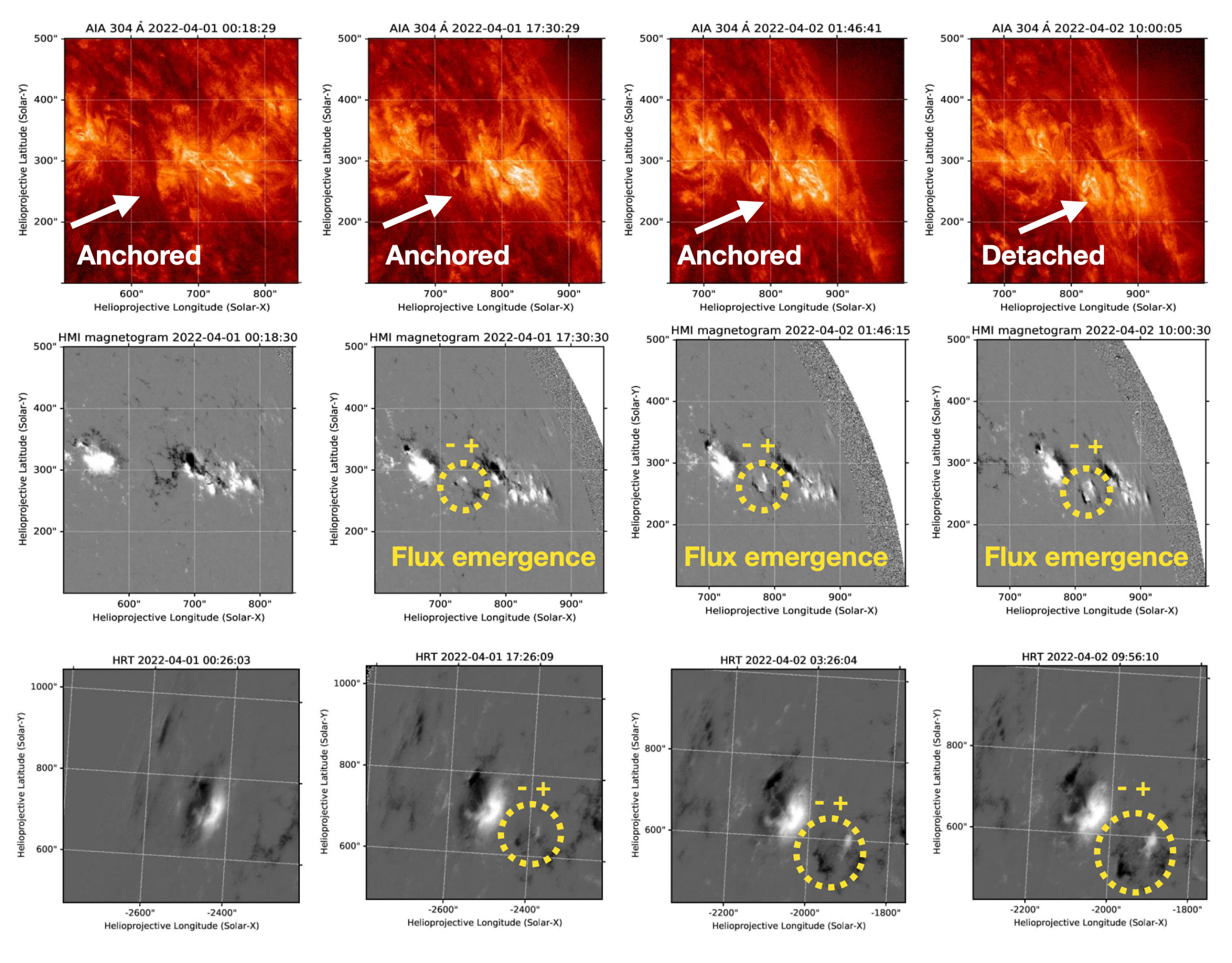}
    \caption{
    SDO/AIA 304 (top row) and HMI (second row), as well as \solo\ PHI/HRT magnetograms (bottom row) taken at different times before the flare. The evolution of the magnetic field constrained with the two different views with SDO/HMI and PHI/HRT shows the emerging bipole (yellow circle, with the `-' sign indicating the negative polarity and `+' the positive one). Since observations are close to the solar limb (western one for AIA and HMI, eastern one for HRT), reversed magnetic polarities appear on the longitudinal magnetic field component on the limb side of true polarities. The emergence of the bipole in time can be seen in the online movie  \url{https://owncloud.ias.u-psud.fr/index.php/s/4iaifYwMjkHSPJ2}.}
    \label{magnetograms}
\end{figure*}

\subsubsection{Hard X-ray imaging spectroscopy with STIX}
\label{stix}

In addition to EUV imaging and spectroscopy, hard X-ray (HXR) imaging spectroscopy is provided for the  M3.9 flare by the Spectrometer/Telescope for Imaging X-rays (STIX) on \solo\ \citep{Krucker2020}. In Fig.~\ref{goes_stix_lc}, we show the soft X-ray fluxes from GOES/XRS together with the background-subtracted STIX HXR count rates in five broad energy bands for the impulsive phase of the flare. The lightcurves show a pronounced pre-flare phase up until $\approx$13:10~UT (time at Solar Orbiter), after which the impulsive phase starts. The HXR emission at the lower energies is mainly caused by thermal plasma, which is reflected by the smooth time evolution. In contrast, the emission above 25\,keV is clearly dominated by non-thermal bremsstrahlung, which shows the typical spiky behaviour which is interpreted as successive episodes of electron acceleration. We note that the non-thermal counts after the third HXR peak are affected by an instrumental issue that will be solved by a new release of the STIX analysis software currently under development. Due to this reason we refrain from performing the full quantitative STIX spectroscopy and will revisit this task in a follow-up paper. Here, we focus on imaging the HXR sources using the STIX pixelated count rates \citep[e.g.][]{Massa2022}.  The results of HXR imaging are discussed in Sect.~\ref{sec-Sources}. 

\begin{figure}
    \centering
    \includegraphics[width=0.51\textwidth]{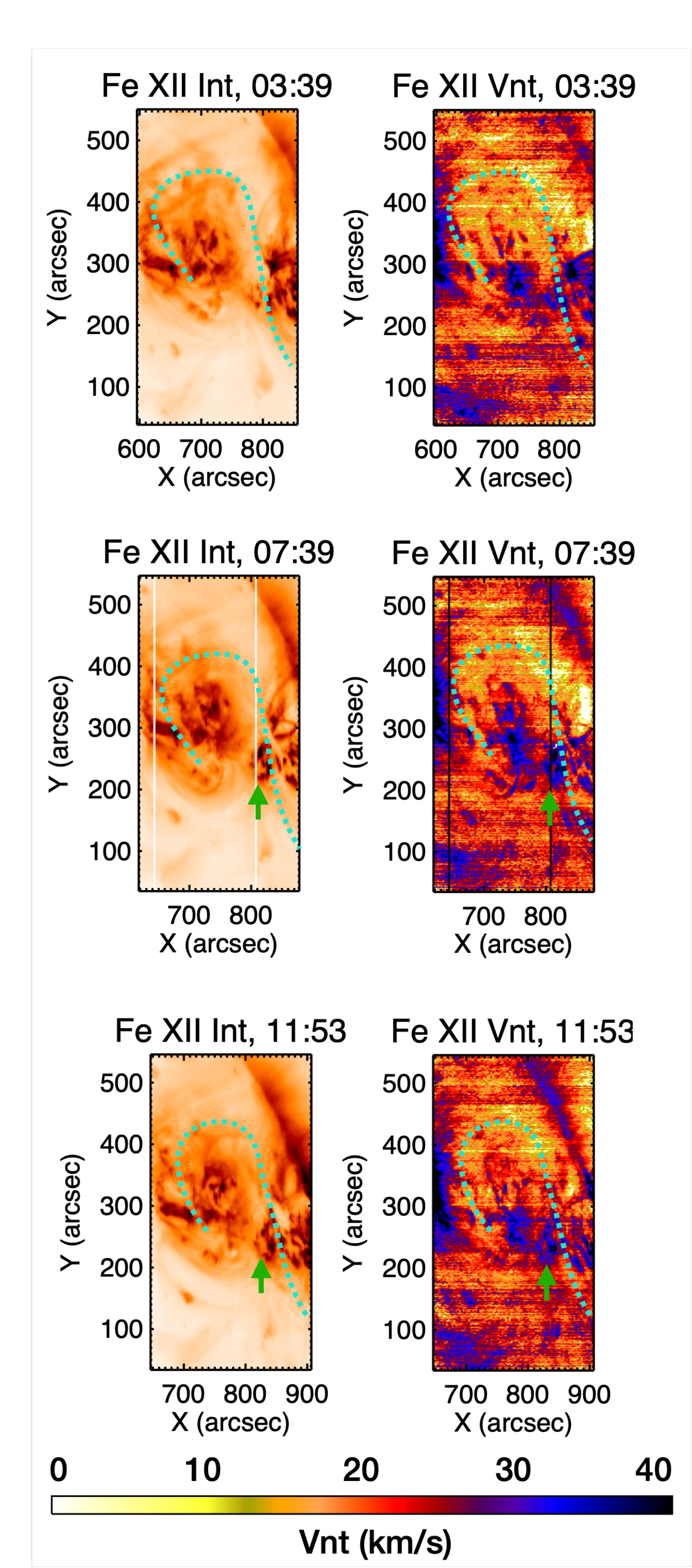}
    \caption{Observations from the Hinode EIS instrument in the pre-flare period.
    Left column: Fe XII intensity maps of the active region over 3 times using data from Hinode EIS. The rasters start at 03:39, 07:39 and 11:53~UT from top to bottom. The colour table is reversed colour, so dark is the highest intensity. The filament is seen as white (low intensity). Right column: Non-thermal velocity of the Fe XII emission line for the same raster times. The filament region (indicated with {cyan dotted lines}) shows enhanced non-thermal velocity in each raster, as well as the emerging bipole (green arrows). The colour bar shows the ranges of the non-thermal velocity in \kms. 
    }
    \label{vnt-eis}
\end{figure}

\section{Precursor activity: From flux emergence to overlying coronal arcade motion during the early filament rise}
\label{sec-Precursor}

\subsection{Flux emergence of a parasitic bipole}
\label{sec-Flux_emergence}

Prior to the eruption, the filament was located between the leading (positive) spot of active region 12976 and the following (dispersed negative) spot of active region 12975, and small negative flux further to the south (Fig.~\ref{overview-activeregions}). The polarity inversion line (PIL) along which the filament resided was the site of supergranular and moat flows during its disk passage, which led to minor flux cancellation in the days leading up to the CME on 2 April.
The reverse-S shaped filament was present when the region rotated over the eastern limb from Earth viewpoint. 

The most significant evolution of the PIL is due to a flux emergence episode that begins on 1 April at 11:30\,UT (yellow dashed circle in Fig.~\ref{magnetograms}). This flux emergence is observed both from SDO/HMI (second row), and from \solo\ PHI/HRT (third row). The bipole emerges on the PIL of the filament, under the filament's axis as can be seen by comparing the position of the new bipole {(second row)} and the SDO/AIA 304 images (top row). The new loops of the emerging flux therefore form and grow under the filament. The leading (following) polarity of the new bipole is positive (negative), see third row, opposite to the orientation of the polarities at the PIL along which the filament resides. 

The bipole emergence implies the formation of a magnetic configuration with a potentially very complex set of interlinking field lines and QSLs, where strong electric current layers are expected to form \citep{Demoulin1996, Aulanier2005}. The emerging bipole further enhances the formation of this current layer by pressing the emerging magnetic field against the pre-existing one above.  When the current layer becomes thin enough, magnetic reconnection sets in and transforms part of the magnetic field supporting the filament to new field connections.  The exact process depends on parameters such as the emerging bipole orientation, its field strength and its location relative to the main PIL \citep{Dacie2018}.  This leads to a progressive detachment of the filament from the lower layers. This could be related to what is observed in the studied case  (as indicated by the white arrow, ``detached" in Fig.~\ref{magnetograms}, {top right panel}). However, to discriminate with certainty other possible interactions between the emerging bipole and the flux rope, or surrounding arcades, a thorough magnetic modelling of the region is necessary and out of the scope of the present paper.


Over the next $\approx$24~h, more and more of the filament appears to become detached from the lower atmosphere and the filament structure becomes more semi-circular {by $\sim$13:20\,UT on 2 April as seen from the AIA perspective}. After sufficient time, such a process typically destabilises the original magnetic configuration.  Then, the filament starts to rise.

In conclusion, the role of the flux emergence that takes place along the PIL could be significant for transitioning the flux rope from a bald patch structure (an {expected configuration during} filament formation via flux cancellation along the PIL) to HFT (slow rise phase/precursor phase), {as present in MHD simulations \citep{Aulanier2010,Zuccarello2015}.} 

\begin{figure*}
    \centering
    \includegraphics[width=\textwidth]{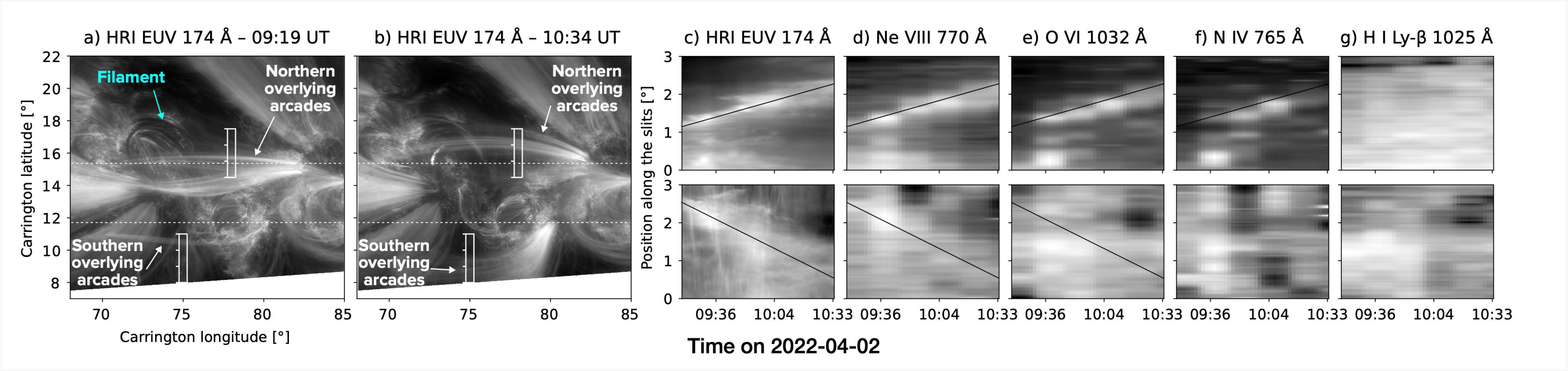}
    \caption{\solo\ EUI/HRI views of the region in between AR 12795 and 12796 at 09:19:15\,UT (a) and 10:34:05\,UT (b). The $x$ and $y$ axes represent respectively the longitude and latitude in Carrington coordinates and are given in degrees. The inverted S-shaped filament in absorption and the overlying coronal loops in emission can be seen in between the two active regions. The two solid white rectangles
    represent the areas where the time stackplots in (c)--(g) have been made.
    The stackplots show the intensity along the slits (northern slit in the top row, and southern slit in the bottom row), for all the frames obtained by EUI/HRI with a cadence of 10\,s (c) and by SPICE with a cadence of 15\,min (d)--(g).
    Four spectral lines observed by SPICE are shown, ordered by decreasing temperature: \ion{Ne}{viii} 770\,\AA~(d), \ion{O}{vi} 1032\,\AA~(e), \ion{N}{iv} 765\,\AA~(f), and \ion{H-Ly}{$\beta$} 1025\,\AA~(g).
    The black solid lines indicate the motion of the northern and southern loops characterised visually for each, and are used to obtain the projected motion speed.
    The temporal evolution of the EUI/HRI and SPICE maps can be seen in the online movie.
    \url{https://owncloud.ias.u-psud.fr/index.php/s/z2eWkHD3gQrHFaK}
    }
    \label{hri-spice-loops-stackplots}
\end{figure*}

\begin{figure*}
    \centering
    \includegraphics[width=\textwidth]{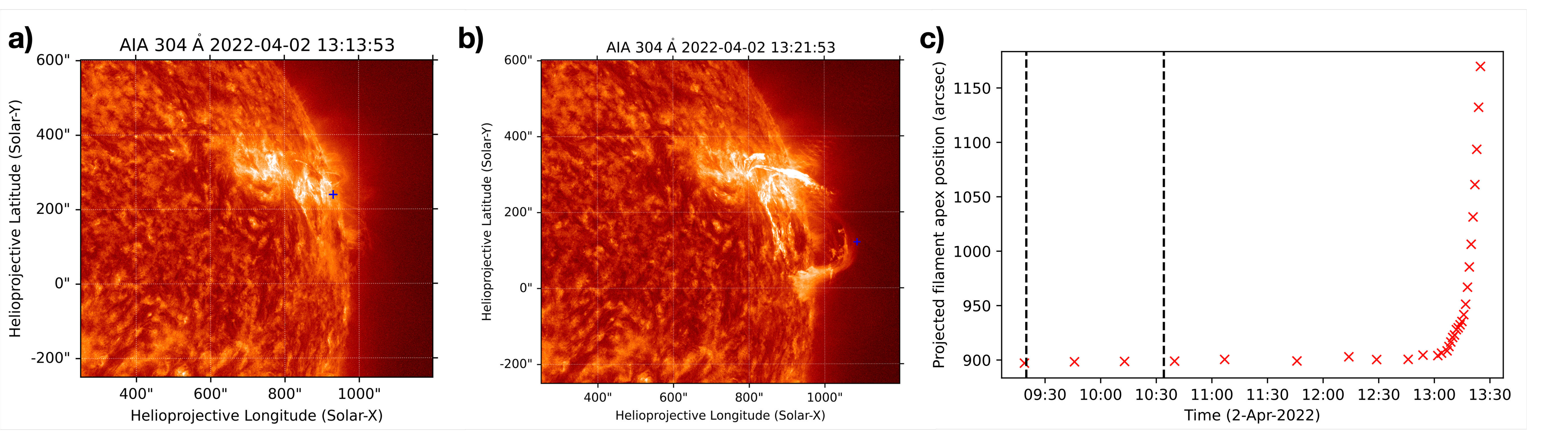}
    \caption{SDO/AIA 304\,\AA\ images of the erupting filament taken at 13:13:53\,UT (a) and 13:21:53\,UT (b) on 2 April. The blue cross indicates the top of the filament used to track the eruption. Each distance from the disk centre 
    (in arcsec) is reported against time in panel c, after removing the solar rotation (see text). The two vertical black dashed lines indicate the start and the end of the EUI/HRI observation window.
    }
    \label{aia-looptop}
\end{figure*}

The intensity and the non-thermal velocities derived from three of the EIS HPW021 rasters are shown in Fig.~\ref{vnt-eis}.
The filament is seen as a dark absorption feature in coronal emission lines in the first three rasters (with reverse colour) but is absent in the fourth raster (not shown) that was taken by EIS during the decay phase of the second M-class flare, which peaked at 17:45~UT. Some of the brightest parts of the active region show the strongest non-thermal velocities. 
The non-thermal velocities around the filament 
are distinctly enhanced especially in the second and third raster.  The second raster at 07:39\,UT shows a clear enhancement of non-thermal velocity at the location of the filament. This demonstrates that, at this time, 5~h before the flare began, the corona is dynamic or turbulent around the region of the filament. This is consistent with the filament detaching from the lower atmosphere and reconnection taking place.  Enhanced non-thermal velocities are an expected consequence of the reconnection from the emergence of the parasitic polarity, as it leads to complex flows. 

\begin{figure*}
    \centering
    \includegraphics[width=1\textwidth]{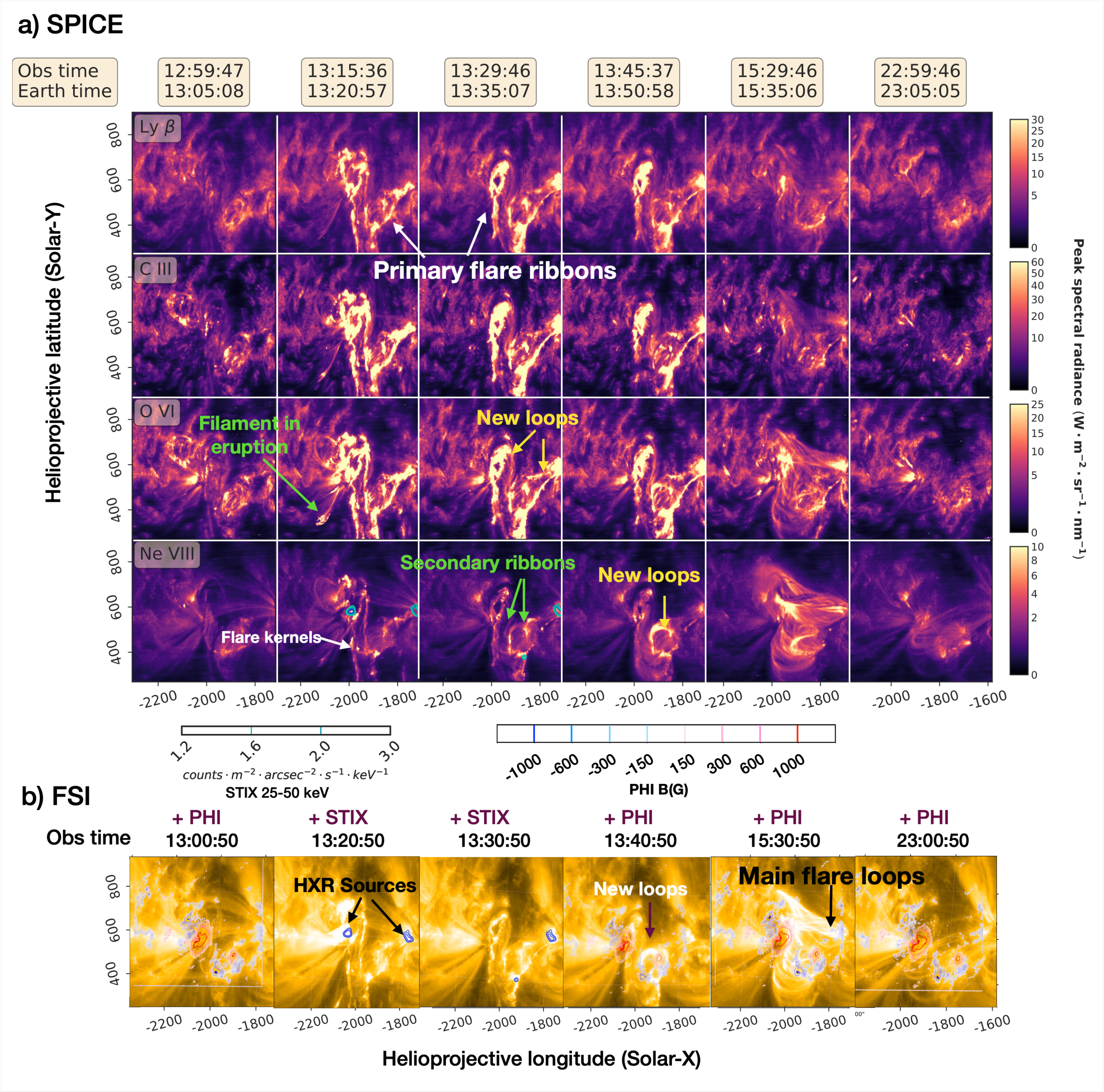}
    \caption{Sequence of observations taken by four different Solar Orbiter remote-sensing instruments. a) Intensity maps for the four ions \ion{H-Ly}{$\beta$} ($10^4$\,K) , \ion{C}{iii} ($10^{4.5}$\,K), \ion{O}{vi} ($10^{5.5}$\,K) and \ion{Ne}{viii} ($10^{5.8}$\,K) {(using the line peak intensity)} and for six different times from 2 April 13:00\,UT until 23:00\,UT. The different structures referenced in the text are indicated with their respective arrow and label. An online movie is available at \url{https://owncloud.ias.u-psud.fr/index.php/s/aESgKF6FXTy5YS5}. 
    b) FSI maps with the same field of view 
    for the closest corresponding times. These maps are overlaid 
    with the closest magnetograms from PHI (contour plots, with red indicating positive polarities and blue negative ones), as well as the non-thermal HXR sources from STIX at 13:20 and 13:30\,UT (contour plots in gardients of green/blue). An online movie is available too at \url{https://owncloud.ias.u-psud.fr/index.php/s/nedpLpAzsmRTdYz}.}
    \label{spice-intensities}
\end{figure*}

\subsection{Overlying coronal arcades}
\label{sec-Overlying}

The EUI/HRI FOV 
was centred on the region between AR 12975 and 12976, therefore capturing the dynamics  of the filament and surrounding coronal loops in the 174\,\AA\ passband (Fig.~\ref{hri-spice-loops-stackplots}) between 09:19:15\,UT and 10:34:05\,UT.

As shown in the two snapshots (a) and (b) in Fig.~\ref{hri-spice-loops-stackplots} and the online movie, a striking apparent motion of both the northern and southern coronal arcades (indicated with white arrows) overlying the filament is present during the observation sequence.  
These loops are anchored in the same western footpoints (their eastern counterparts are not visible). This fixed position is indicated as the horizontal dashed white lines in Fig.~\ref{hri-spice-loops-stackplots}a,b.
By choosing two areas across the northern and southern loops where the intensity of the loops is best seen during the EUI/HRI observation time, we investigated the motion of the bright front of these loops in time-distance plots (Fig.~\ref{hri-spice-loops-stackplots}c) defined on EUI/HRI images (Fig. \ref{hri-spice-loops-stackplots}a,b). By visualising the intensity along the synthetic slits ($y$-axis) over time ($x$-axis), the highest intensity regions are seen to move monotonically, giving an apparent motion speed of $v_{\textrm{north}}=3.1$ \kms. {The motion of the southern loops is less smooth, which renders a less clear linear fit with an apparent speed of} $v_{\textrm{south}}=5.5$ \kms.  This motion characterised visually is indicated with the black solid lines.
This apparent motion could either be due to an actual motion of the loop tops or a change in the density or temperature so that loops appear subsequently (similar to the sequential evaporating plasma filling within flare loops, which consequently appear as growing in height). 
To test these two scenarios, we used the SPICE dynamics studies that were running about every 15\,min in the same time frame. 
In Fig.~\ref{hri-spice-loops-stackplots} d--g, time-distance plots are shown for different ions where the overlying arcades are seen.
These time-distance plots were constructed from the intensity maps obtained via the Gaussian fitting described in Sect. \ref{spicedata}.
For the northern loop, the motion seen in EUI/HRI EUV (Fig.~\ref{hri-spice-loops-stackplots}c) is also visible in the SPICE lines \ion{Ne}{viii} 770\,\AA, \ion{O}{vi} 1032\,\AA, and \ion{N}{iv} 765\,\AA, with peak temperature formation at  0.63, 0.32, and 0.13\,MK, respectively. The dark solid lines indicate this motion found for each of the ions.
For the southern loop, the motion is harder to identify in SPICE because of the lower contrast, and is only visible in lines \ion{Ne}{viii} 770\,\AA~and \ion{O}{vi} 1032\,\AA.
No sign of the motion was found in \ion{H}{i} \ion{Ly}{$\beta$} 1025\,\AA, suggesting a temperature higher than 0.01\,MK. To conclude, the associated EUI/HRI and SPICE stackplots show that the apparent motion of the loops corresponds to a multi-thermal structure that is moving in one block, rather than a heating and cooling process.

\begin{figure*}
    \centering
    \includegraphics[width=1\textwidth]{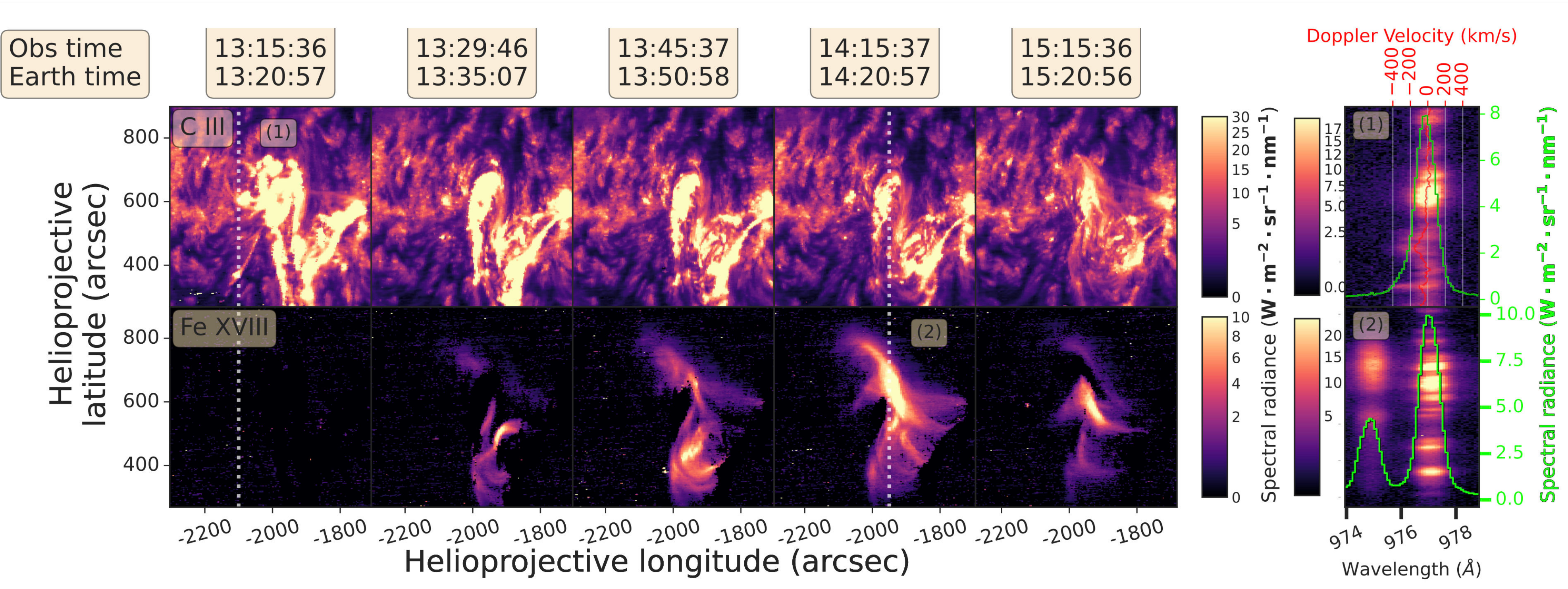}  
    \caption{Intensity maps for the \ion{C}{iii} ($10^{4.5}$\,K) and \ion{Fe}{xviii} ($10^{6.9}$\,K) lines {(using the line peak intensity)} for five different times around the peak of the flare. The dashed lines, {in the first and fourth panels,} correspond to the positions and times of the two spectograms shown on the right. Spectrogram (1): data from raster starting at 13:15:35~UT located around -2100\,\arcsec\ in helioprojective longitude, showing the emission peak of \ion{C}{iii} and the absence of \ion{Fe}{xviii} {(interpolated over their positions)}.
    Spectrogram (2): data from raster starting at 14:15:37~UT located around -1950\,\arcsec\ in helioprojective longitude, showing the emission peaks of \ion{C}{iii} and \ion{Fe}{xviii} {(interpolated over their positions)}.
    The green curve is the mean spectrum over all selected pixels in latitude in the spectrogram while the red curve indicates the wavelength position of the peak line intensity.
    The width of the spectral profiles is dominated by the instrumental broadening.
    The Doppler velocity scale for \ion{C}{iii} is written in red at the top. 
    }
    \label{spice-fexviiimaps}
\end{figure*}

\subsection{Comparison with the standard model in 3D}
\label{sec-Comparison}

This apparent motion of overlying arcades is expected from the 3D standard model. In the 10 September 2014 X-class flare analysed by \citet{Dudik2014}, the authors showed a growing system of expanding warm coronal flare loops in the precursor phase and on the active region periphery (Fig.~2 in that paper). These overlying loops had an apparent expansion motion ranging from $\approx 6$\,\kms\ to 21\,\kms\ in the AIA 171\,\AA\ images (their Fig.~3), similar to that found in the EUI/HRI data here (Fig.~\ref{hri-spice-loops-stackplots}). Furthermore, they were seen to move co-temporal to the start of the filament rise. In their Fig.~11, \citet{Dudik2014} showed a numerical modelling of the 3D standard flare model made with the OHM code \citep[see details in][]{Aulanier2010, Aulanier2012, Janvier2013}.  In particular, the authors pinpointed a series of overlying loops pushed by the flux rope expansion. They explained that the overlying coronal loops expand and move sideways due to the unstable flux rope pushing them as the expansion starts. Overall, the expanding behaviour observed in their paper and obtained here with the highest spatial resolution and temporal resolution with EUI/HRI are consistent{, pointing out that the observations are well within the predictions given by the standard flare model in 3D.} 

Next, since the filament is seen as late as 13:30~UT in the AIA FOV (ultimately as a prominence eruption), we report in Fig.~\ref{aia-looptop} the position of the top part of the filament against time. This position is corrected from the solar rotation by taking into account the actual rotation speed of the region on the solar disk starting from 25 March.
The EUI/HRI observation time window is indicated with the vertical dashed lines in panel (c). During this time, the filament top is not seen to rise significantly. This profile is similar to the ones studied in the filament early rise analysis of \cite{Cheng2020} in stackplots of intensity vs. time as a measure of height evolution in time. They found that the early, slow rise phase preceding the main filament acceleration phase is a common feature, with both having different dominant mechanisms. Here, the early rise phase has been shown to be linked to the parasitic bipolar emergence.

\section{Eruption dynamics in multi-thermal observations}
\label{sec-Eruption}

\subsection{Multi-thermal flare ribbons and flare loops seen by SPICE}
\label{sec-SPICE}

We show in Fig.~\ref{spice-intensities}a intensity maps obtained from the intensity peak parameter of the line Gaussian fits. These maps show the spectral radiance of the peak of emission for four of the SPICE emission lines. Each row is ordered by the emission temperature of the species (see Table~\ref{table-Ions}), starting with \ion{H-Ly}{$\beta$} at the top and \ion{Ne}{viii} at the bottom. The columns correspond to different raster times, from before the flare to ten hours after. The start time of each raster is given as observation time at the spacecraft (`Obs time') and the projected time at the Earth (`Earth time').

The SPICE raster that corresponds to the filament eruption began at 13:15:36\,UT (second column) and was completed at 13:29:46\,UT at the spacecraft, and runs from right to left. These correspond to times 13:20:57 and 13:35:07\,UT at Earth. 
Figure~\ref{spice-intensities}b presents the maps of the same region obtained with EUI/FSI 174\,\AA. These maps are overlaid with the closest in time magnetograms obtained by  PHI/HRT (LOS magnetic flux density contours from 13:00, 13:40, 15:30 and 23:00\,UT), as well as HXR from STIX showing the non-thermal HXR sources (intensity contours between 25-50 keV) at 13:20 and 13:30\,UT (see also Sect. \ref{sec-Sources} for a description of this dataset).

Flare ribbons are seen clearly as high intensity emissions in three of the lines (\ion{Ly}{$\beta$}, \ion{C}{iii}, and \ion{O}{vi}), while the \ion{Ne}{viii} line shows some of the most intense flare kernels. The system of ribbons is more complex than the typical {two} $J$-shape ribbons often reported in the literature {in the context of eruptive flares} \citep[see, e.g.][and references therein]{Moore1995,Demoulin1996b,Chen2011}. 
We distinguish primary flare ribbons as indicated with the white arrows, with the eastern arrow showing a hooked, J-shaped structure  expected for a flux rope eruption \citep[see e.g.][]{Demoulin1996b,Janvier2014}. As the flux rope spine extends in the N-S direction, the western ribbon extends southwards outside of the instrument's FOV.  

Next, we observe a set of two small flare ribbons seen to appear at 13:15\,UT (indicated with green arrows in the 13:29\,UT frame {of \ion{Ne}{viii} line}). The EUI/FSI 174\,\AA +  PHI/HRT overlay obtained at 13:40\,UT shows that this area also corresponds to a secondary set of flare loops, which are anchored in the emerging bipole region described in Fig.~\ref{magnetograms}. This second set of flare ribbons and new loops correspond to a complex magnetic field configuration and associated reconnection: see Sect. \ref{sec-Flux_emergence}. 
The EUI/FSI 174\,\AA + PHI/HRT overlay shows that flare loops are both anchored between the positive polarity of AR 12976 and the negative polarity of the emerging bipole, {as well as that of the positive polarity of the emerging bipole and} the negative polarity of AR 12975. The different orientations of the flare loops are also visible in the hotter \ion{Ne}{viii} line.


We also observed the \ion{Fe}{xviii} 974.86\,\AA\ emission line in the SPICE dataset. This line has been observed previously \citep[e.g.][]{2012ApJ...754L..40T} in active regions with the Solar Ultraviolet Measurements of Emitted Radiation \citep[SUMER:][]{1995SoPh..162..189W} experiment. 
The \ion{Fe}{xviii} line lies close to the strong \ion{C}{iii} 977\,\AA\ line and the lines were fitted together with a double-Gaussian fit. Two examples of these fits are shown in Fig.~\ref{spice-fexviiimaps}. In this figure, we also show the intensity maps for these two lines for five rasters, corresponding to times before and after the peak flare emission. No emission from the \ion{Fe}{xviii} line is seen in the raster commencing 13:15:36\,UT,  
while a set of flare loops are seen in the following raster starting at 13:29:46\,UT. 
These flare loops correspond to the same set of secondary flare loops seen in the 13:40\,UT EUI/FSI 174\,\AA + PHI/HRT overlay shown in Fig.~\ref{spice-intensities}b and anchored {on one side} in the emerging bipole. 
Further hot flare loops are seen to appear subsequently, while the \ion{C}{iii} intensity map shows the flare ribbons at the footpoints of the flare loops (Fig.~\ref{spice-fexviiimaps}). We notice that the second set of flare loops to appear at 13:45\,UT is connecting the positive polarity of AR 12976 to the negative polarity of the emerging bipole, as discussed above for Fig. \ref{spice-intensities}. 

{In summary, the} different emission lines obtained with SPICE show a complex set of flare ribbons and loops due to the magnetic field configuration (main bipole and emerging parasitic bipole). 
In the following, we investigate the Doppler shifts obtained with SPICE as well as with EIS.

\begin{figure*}
    \centering
    \includegraphics[width=\textwidth]{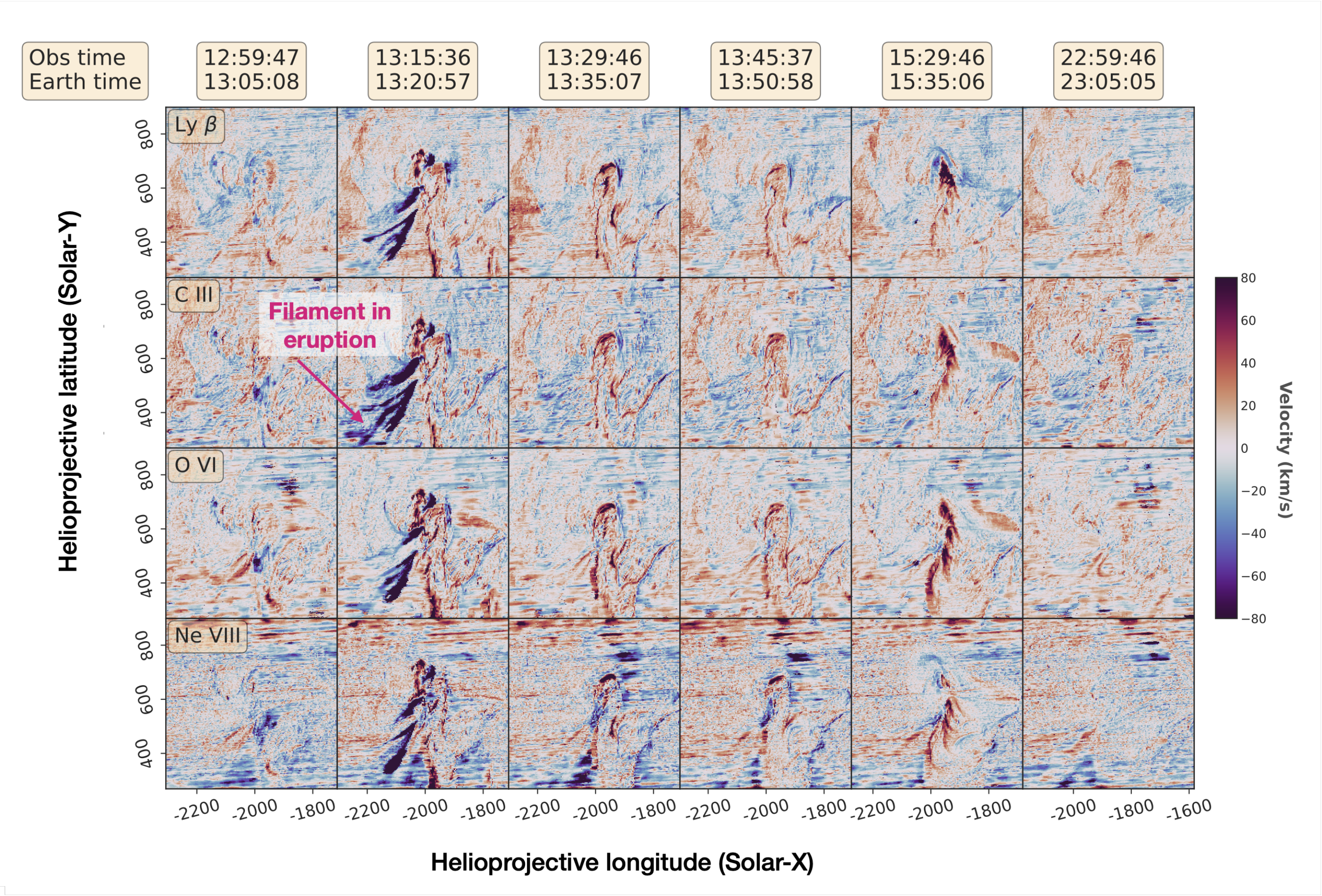}
1    \caption{SPICE Doppler velocity maps for the four ions \ion{Ly}{$\beta$}, \ion{C}{iii}, \ion{O}{vi} and \ion{Ne}{viii} and for five different times from 2 April 13:05\,UT until 3 April {23:05\,UT} 
(Earth time). The corresponding intensities are shown in Fig. \ref{spice-intensities}. The different structures referenced in the text are indicated with their respective arrows and labels. }
    \label{spice-doppler}
\end{figure*}

\subsection{Filament eruption captured by SPICE and EIS}
\label{sec-Filament}

\subsubsection{Erupting plasma as seen by SPICE}
\label{sec-spicedoppler}

The Doppler velocities are analysed in the following from the `dynamics' rasters, and are directly deduced from the spectrum intensity peak shifts. However, the wavelength reference is not precisely calibrated and the flight performance \citep[as discussed in Appendix~B of][]{Fludra2021} also alters the results. Consequently, the wavelength used as a reference was calculated directly from each raster, by considering the mean wavelength value over a selected region considered as quiet Sun region, and the Doppler velocities given further below are relative to this reference wavelength rather than absolute.

For the same spectral lines and raster scans as in Fig.~\ref{spice-intensities}, we show the corresponding Doppler velocity maps in Fig.~\ref{spice-doppler}. In the raster commencing at 13:15:46\,UT,  
the strong blue shifted components are clear, with velocities reaching up to 400\,\kms\ (see e.g. spectrogram (1) {on the right of} Fig. \ref{spice-fexviiimaps}). The multi-stranded appearance of the structure moving upwards is reminiscent of the filament leg seen in the AIA and EIS frames and discussed further below (see {{Figs.}} \ref{fig.movie1} and \ref{fig.eis-aia}). While most of the filament structure corresponds to a low intensity area in the different lines in Fig.~\ref{spice-intensities}, one can however see some strong emission in a collimated thin region indicated with the green arrow {in the third row of Fig.~\ref{spice-intensities}}. 

Furthermore, a cross-section of the SPICE data cube has been selected in Fig.~\ref{spice-fexviiimaps} (dashed grey line in (1) {on the left image}) which relates to the moving filament leg. High values ($> 100$\,\kms) of Doppler shifts are seen there, as indicated by the spectrogram of $I(\lambda,y)$ to the right (spectrogram 1). 
We also observe red shifted line emission along the flare ribbons indicated with black arrows in Fig.~\ref{spice-doppler}, and some strong red shift along the flare loops, as in the raster commencing at 15:29:46\,UT.

In the SPICE data, there is however some uncertainty in the Doppler shifts in that intensity gradients will cause red and blueshifts in the direction along the slit next to bright features. This is because the point spread function (PSF) of SPICE is elliptical and rotated. While first steps have been taken to correct this effect \citep{Plowman}, here we use the original SPICE data. Still, because the flare ribbons are associated with only redshifts (and no blueshifts), it seems reasonable to assume that the redshifts are real and not simply due to the SPICE PSF. To help differentiate what features might be due to the astigmatism and what may be solar in origin, such as the strong Doppler shift values found related to the filament leg, we extend the analysis further in Appendix \ref{spice-astigmatism}. Following this further step, we ensured that that the strong blueshifts are indeed of solar origin, and that the redshifts indeed indicate downflows.


\subsubsection{Blue-shifted component in the filament leg as seen by EIS}
\label{sec-eisdoppler}

\begin{figure*}
    \centering
    \includegraphics[width=0.8\textwidth]{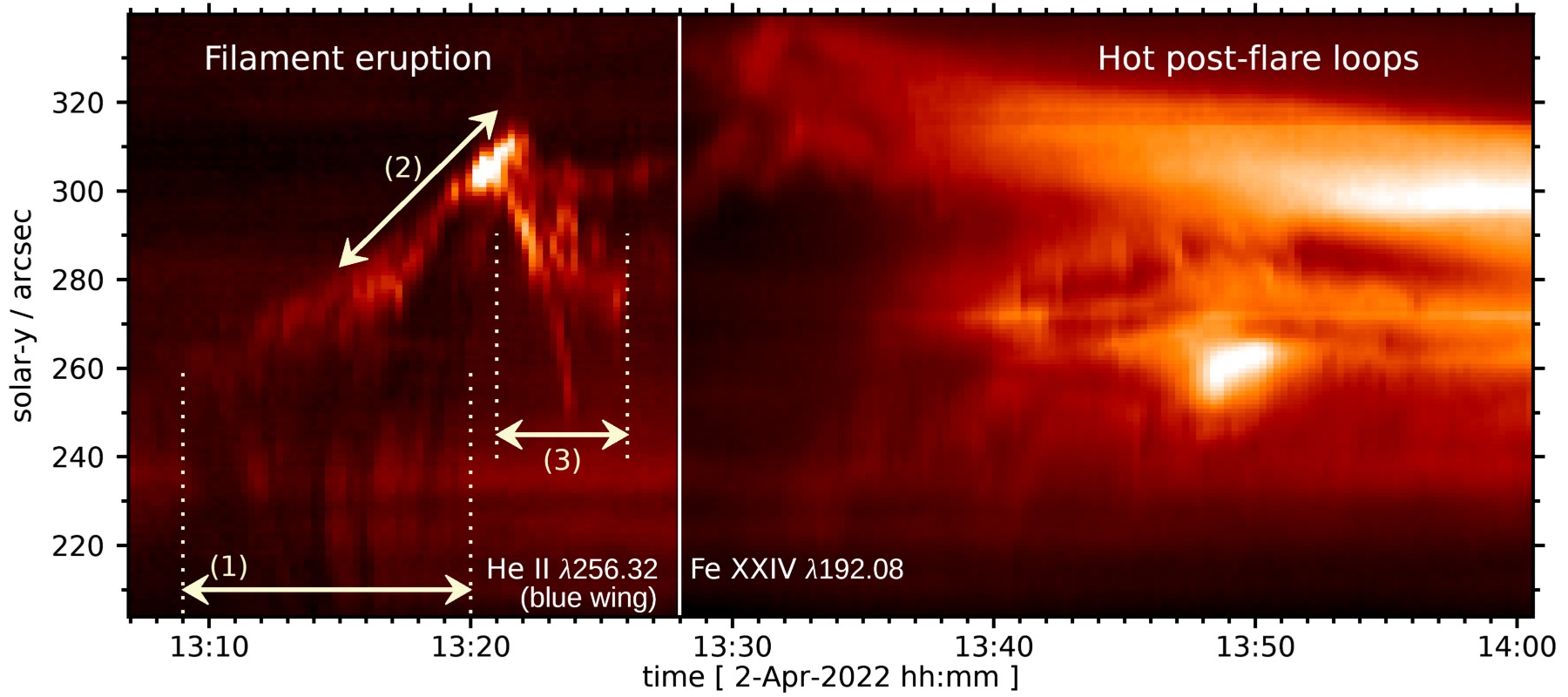}
    \caption{Hinode/EIS observations over the flaring period. The left panel of this image shows the intensity of the \ion{He}{ii} 256.32\,\AA\ line observed by EIS, as a function of time and solar-$y$ position, formed in the blue wing of the line between $-190$ and $-10$~\kms.  Three features discussed in Sect. \ref{sec-eisdoppler} 
    are highlighted with arrows and numbers. The right panel shows the intensity of the \ion{Fe}{xxiv} 192.03\,\AA\ line. A square-root intensity scaling has been applied for both lines.}
    \label{fig.composite}
\end{figure*}

\begin{figure*}
    \centering
    \includegraphics[width=\textwidth]{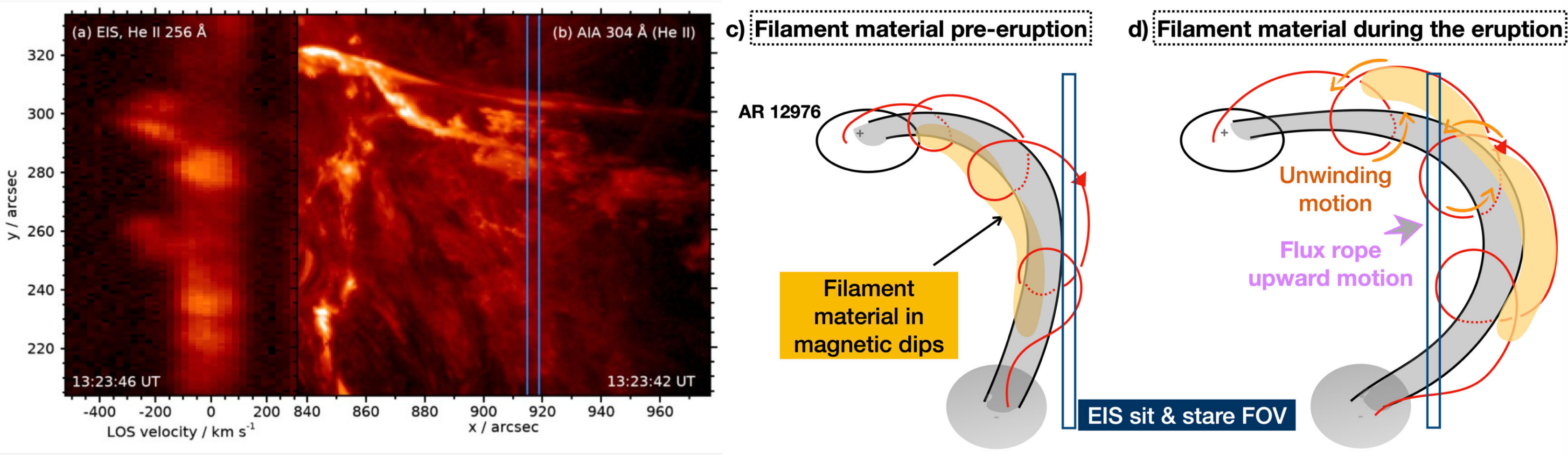}
    \caption{A focus on the filament eruption as seen from Hinode/EIS. (a) A frame from a movie showing the evolution of the EIS \ion{He}{ii} 256.32\,\AA\ line, and (b) the AIA 304\,\AA\ emission at 13:23~UT. A square-root intensity scaling is used for the EIS images. Two vertical blue lines on the AIA images show the location of the EIS slit. The online movie (between 13:07~UT to 13:27~UT) is available at \url{https://owncloud.ias.u-psud.fr/index.php/s/z86YZGWM9jZtAcL}. (c,d) Diagrams summarising the interpretations of the observations, in particular the unwinding motion of the magnetic flux rope supporting the filament. 
    } 
    \label{fig.movie1}
\end{figure*}

The sit-and-stare EIS observation began prior to the flare impulsive phase and continued through most of the long decay phase. As the slit was located to the west of the flare ribbons, this aspect of the flare evolution could not be studied. However, the filament eruption was observed from 13:10~UT to 13:35~UT, and the cooling of hot post-flare loops was captured from 13:30~UT to the end of the observation at 17:08~UT. Figure~\ref{fig.composite} shows these two aspects of the event as they appear in a continuous time sequence ($x$-axis) of 54~min. The \ion{He}{ii} 256.32\,\AA\ line is used to show the complex intensity variation associated with the filament eruption during the first 21~min.
Next, from 13:28~UT onwards, the intensity in Fig.~\ref{fig.composite}  comes from \ion{Fe}{xxiv} 192.03\,\AA, the hottest line observed by EIS. It shows the development of hot post-flare loops at the spatial locations through which the filament had previously passed. The 192.03\,\AA\ line is principally formed in the temperature range 13 to 33\,MK, and so the presence of this line gives supporting evidence for the high-temperature component found from STIX data (Sect.~\ref{sec-Sources}).

The remaining of this section will focus on aspects of the filament eruption. 
Three features are highlighted on the left panel of Fig.~\ref{fig.composite} from the  filament eruption stage:
\begin{description}
\item[EIS-1:] Between 13:09~UT and 13:20~UT a succession of dark filament threads are seen in absorption against the background \ion{He}{ii} emission, and are approximately parallel to the EIS slit. 
These reveal the slow, early rise phase of the filament.
\item[EIS-2:] A bright section of the north filament leg moves progressively northwards during 13:12~UT to 13:22~UT due to the stretching and straightening of the leg (Fig.~\ref{fig.movie1}). It becomes very intense at 13:21~UT.
\item[EIS-3:] A tangle of bright plasma clumps and twisted threads that appear to unwind from EIS-2 after it reached its maximum brightness.
\end{description}

Figure~\ref{fig.movie1} shows a frame from a movie that compares the evolution of the \ion{He}{ii} emission along the EIS slit to the AIA 304\,\AA\ images, which is dominated by the \ion{He}{ii} 304\,\AA\ emission line. Both datasets were spatially aligned, with details of the method given in Appendix \ref{eis_aia_alignement}.

For the EIS frame (panel a), the $x$-axis is wavelength converted to line-of-sight (LOS) velocity. 
Panel b shows the AIA image nearest in time to the EIS exposures, as determined from the mid-points of the exposures. The $y$-ranges for the two sets of images are the same, and the location of the EIS slit is shown on the AIA image. Inspection of the EIS spectral images shows many transient, compact brightenings on the short-wavelength side of the 256.32\,\AA\ line with LOS velocities between $-50$ and $-400$~\kms. These can be directly associated with bright filament plasma in the AIA images, and demonstrate that the filament plasma is moving towards the Earth. 

For the simplistic model of the filament leg being a curved cylinder (as shown in panels c and d) that is approximately moving radially away from the Sun's centre, the Doppler shifts imply plasma moving away from the Sun. 
There may also be a tilting of this cylinder towards the Earth during the early phase of the eruption that contributes to the blue-shifted emission.

The standard flare model predicts that the flux rope will unwind as the eruption progresses. This was shown, for example, in panels (g, h, i) of Fig.~1 in \citet{Barczynski2020} where the numerical modelling made with the OHM code \citep{Aulanier2012} allowed to quantify the flux rope expansion. 
This MHD simulation confirmed the predictions that the twist per unit length decreases due to this expansion.

Here, evidence for this unwinding comes from both AIA image sequences and the EIS spectral data. Between 13:10~UT and 13:15~UT, faint, thin emission threads can be seen in front of the dark absorption cloud of the filament at about position $(+870,+315)$\as\ in the AIA 304\,\AA\ images (movie attached to Fig.~\ref{fig.movie1}).
They move southwards across the absorption patch and, since the {bright threads are covering the background features,} provided that they are optically thick,  they must be in front of the absorption patch and thus on the Earth-facing side of the filament. The southward motion then implies an anti-clockwise {winding motion (Fig.~\ref{fig.movie1}d).} 
Later, between 13:19~UT and 13:23~UT, this same effect is seen more clearly around position (+900,+305) in the transition from feature EIS-2 to EIS-3 of Fig.~\ref{fig.composite}.

\begin{figure}
    \centering
    \includegraphics[width=0.50\textwidth]{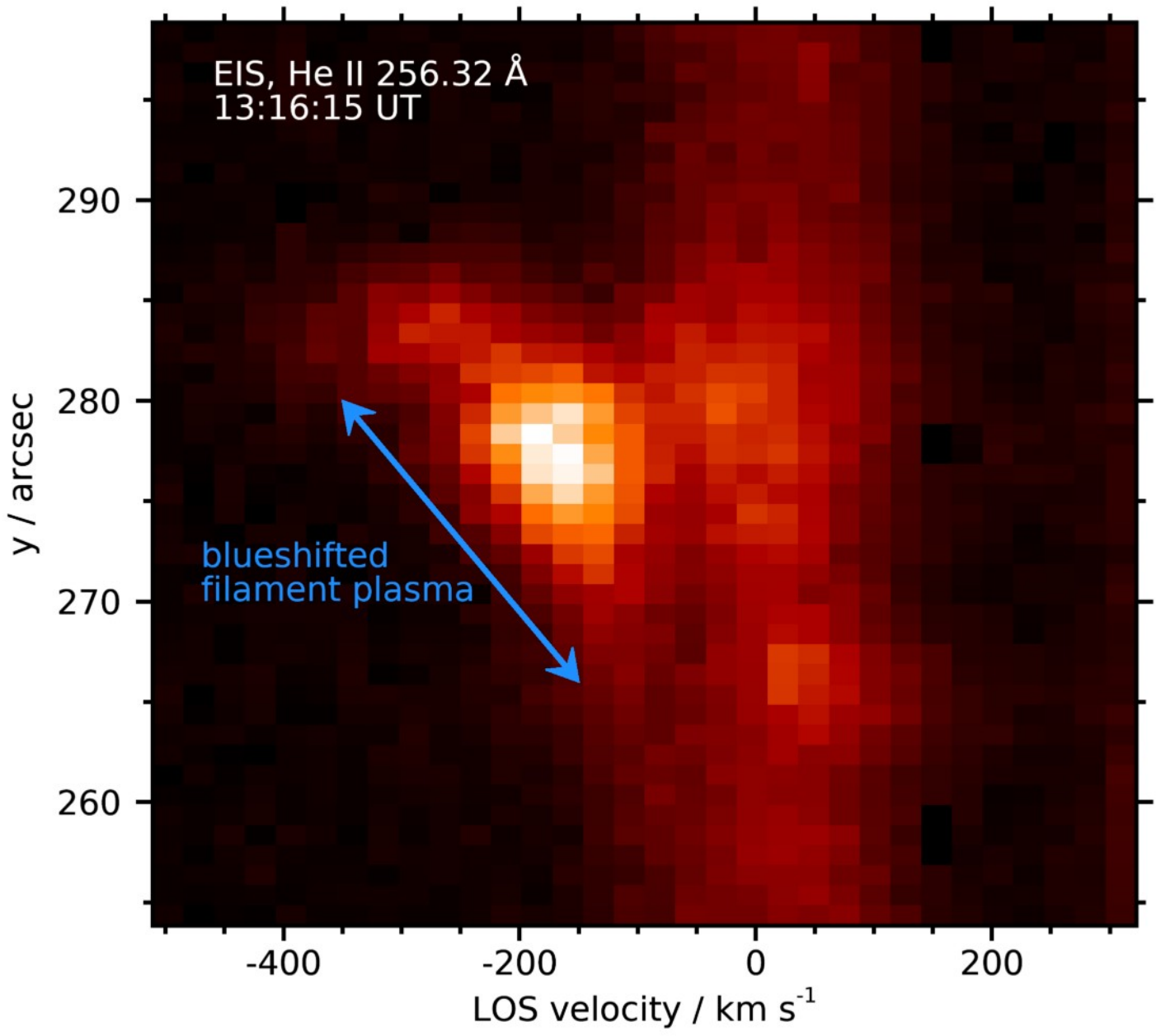}
    \caption{An EIS exposure from 13:16:15~UT showing the variation of the \ion{He}{ii} 256.32\,\AA\ {line} along the slit. A blue arrow indicates the 
    heated filament plasma that shows an increasing blueshift with $y$-position, giving evidence of a clockwise {unwinding} for the filament {(see Fig. \ref{fig.movie1} c,d)}. }
    \label{fig.he2-twist}
\end{figure}

Further evidence for an anti-clockwise motion comes from the Doppler-shifted \ion{He}{ii} 256.32\,\AA\ emission. Fig.~\ref{fig.he2-twist} shows the line profile as a function of $y$-position at 13:16:15~UT. The bright, erupting filament plasma is seen as a ``spur" on the rest wavelength emission that extends along the slit. The plasma becomes more blueshifted with increasing $y$ between $y=265$\as\ and $y=285$\as. If the filament is interpreted as  {embedded in a flux rope} 
(see Fig. \ref{fig.movie1} d)
then an anti-clockwise {unwinding implies that} plasma on the north side of the cylinder moves towards the observer with a greater speed than plasma on the south side. We note that as the filament further evolves and unwinds, then some of the windings will not be able to support the material against gravity anymore, with material being dragged up by the unstable field and with material flowing down along the legs \citep[see e.g.][]{Sasso2014}. Here, the observations are only catching the early unwinding phase.

\begin{figure*}[t]
    \centering
    \includegraphics[width=\textwidth]{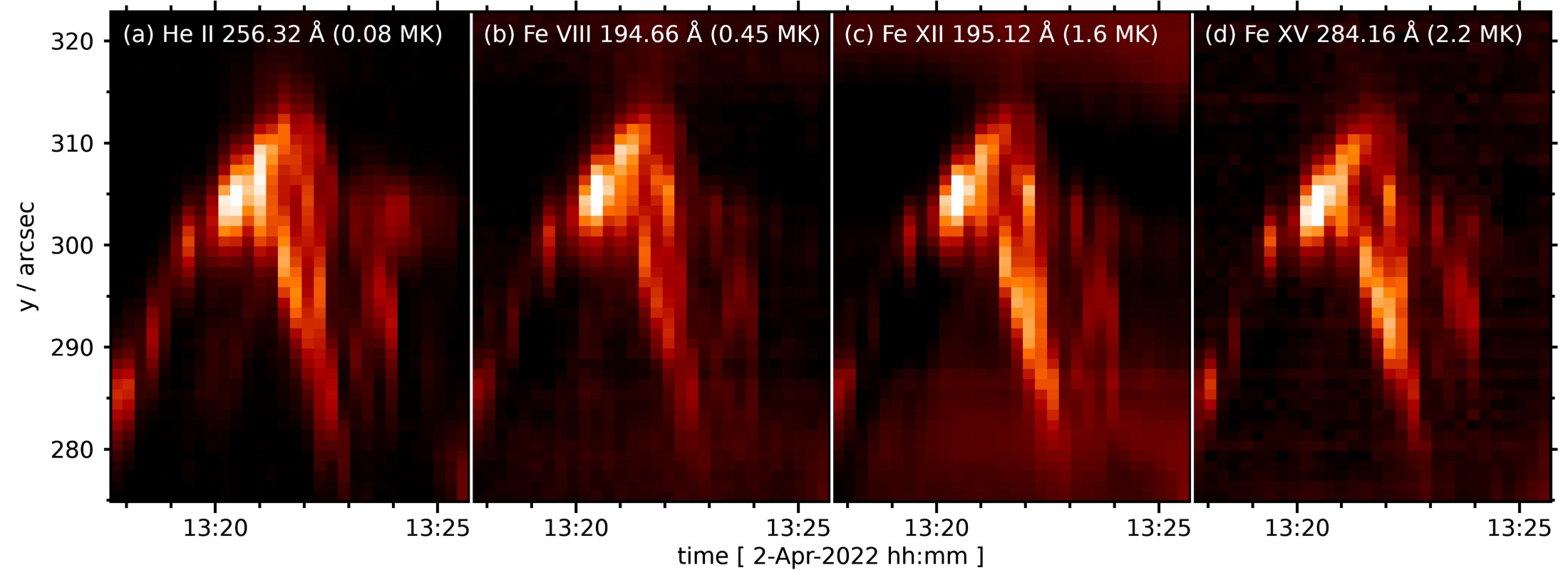}
    \caption{Intensity maps as  functions of time and $y$-distance for four EIS emission lines, showing the transition from feature EIS-2 to feature EIS-3 (Fig.~\ref{fig.composite}). Square-root intensity scaling is used for each image.}
    \label{fig.multitemp}
\end{figure*}

{Next,} Figs.~\ref{fig.composite} and \ref{fig.movie1} show the EIS \ion{He}{ii} emission, which corresponds to plasma at around $8\times 10^4$\,K. The transition from absorption in \ion{He}{ii} to emission implies heating of the initially cold filament plasma as the eruption proceeds. The heating leads to plasma with a wide range of temperature, as demonstrated by Fig.~\ref{fig.multitemp}. This is similar to what is found in both the intensity and Doppler maps obtained from the SPICE dynamics studies with a relatively wide range of temperatures.
The part of the filament evolution that corresponds to the transition from feature EIS-2 to EIS-3 {(Fig.~\ref{fig.composite})} is shown in four different emission lines. The temperatures of maximum ionisation (obtained from the CHIANTI database) are displayed. We conclude that the filament evolution is remarkably similar in all four lines. This demonstrates that each bright clump or thread of the filament is a multi-thermal structure. The filament plasma is weakly visible in \ion{Fe}{xvi} 262.99\,\AA\ (2.8\,MK), but not in hotter lines. In particular, the filament is not heated to flare temperatures (10\,MK).

\begin{figure*}
    \centering
    \includegraphics[width=\textwidth]{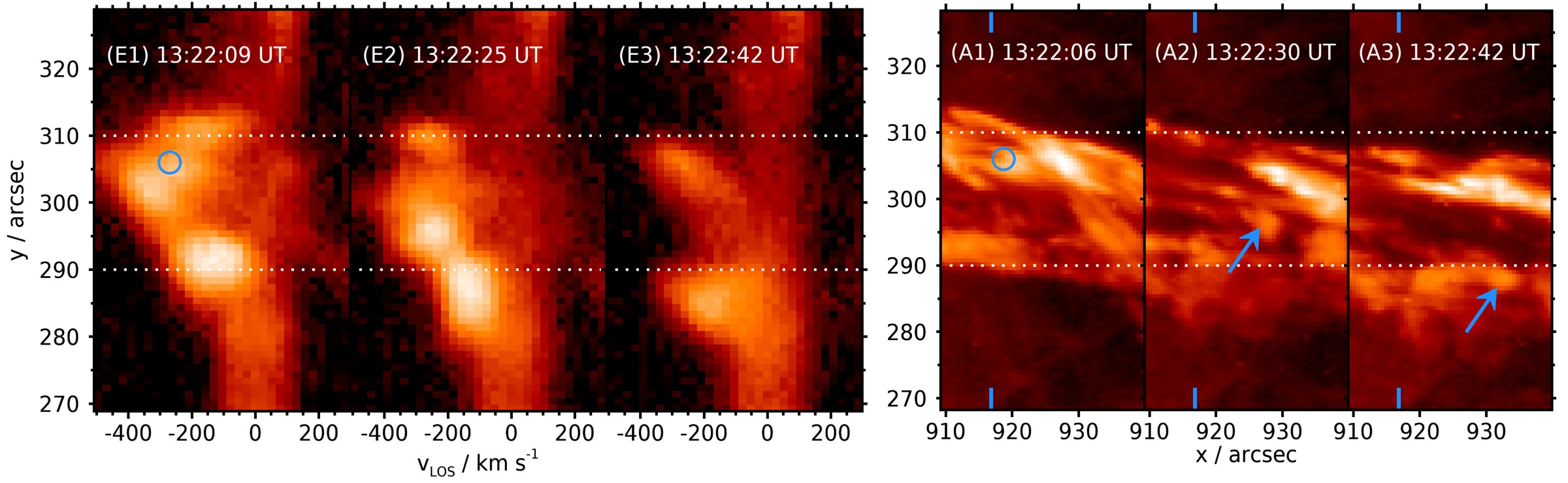}
    \caption{Comparisons betwen Hinode/EIS and SDO/AIA observations of the filament material. Panels E1--E3 show three consecutive EIS exposures of \ion{He}{ii} 256.32\,\AA\ {line}. Panels A1--A3 show nearest-in-time AIA 304\,\AA\ images with the same $y$-range as panels E1--E3. The location of the EIS slit on the AIA images is indicated with two short vertical blue lines {at the panel top and bottom}. A logarithmic intensity scaling is used for all images. The horizontal dotted lines are used to help compare different panels. Features marked with circles and arrows are discussed in {Sect. \ref{sec-eisdoppler}.} 
    }
    \label{fig.eis-aia}
\end{figure*}

As feature EIS-3 evolves, a number of fast-moving plasma clumps are seen in the AIA 304\,\AA\ images and so there is the opportunity to compare LOS velocities from EIS with plane-of-sky (POS) velocities from feature-tracking of the AIA clumps. Fig.~\ref{fig.eis-aia} shows three consecutive \ion{He}{ii} 256.32\,\AA\ exposures from EIS (panels E1--E3). Emission from the solar surface behind the filament gives rise to fairly uniform emission along the length of the slit close to the rest wavelength of the line. A number of bright, compact features are seen on the short-wavelength side of the profile that belong to the filament. The centroids of the features range in LOS velocity from $-140$~\kms to $-340$~\kms, which translate to radial speeds of 340~\kms\ to 820~\kms (see Appendix \ref{los_radial_speeds}).  

Panels A1--A3 {of Fig.~\ref{fig.eis-aia}} show the three AIA 304\,\AA\ images that are closest in time to panels E1--E3. The spatial coverage in $y$ is the same for all panels, and the horizontal dotted lines can be used to help compare features between panels. The location of the EIS slit on an AIA image is indicated by short blue lines at the top and bottom of the image. Arrows on panels A2 and A3 highlight a plasma clump that can be tracked for three frames (it is also seen at 13:22:18~UT). The motion of this clump corresponds to a POS speed of 520~\kms, which translates to a  radial speed of 580~\kms. The clump locations are to the right of the EIS slit. We estimate that the clump was at the EIS slit position for the 13:22:09~UT exposure (panel E1). The circle on panel A1 shows the estimated position of the clump in the AIA 13:22:06~UT exposure based on extrapolation, although the clump itself cannot be clearly identified in this exposure. The corresponding position on the EIS slit is indicated by the circle on panel E1, and there is emission here with a centroid at $-270$~\kms. This translates to a radial speed of 650~\kms, which is larger than the speed derived from AIA but not inconsistent given the uncertainties involved.

In summary, the EIS data provide high time resolution spectral data of the filament eruption and the development of post-flare loops. The filament shows {unwinding} motions in both the EIS and AIA data, consistent with 3D models of solar eruptive events. Filament plasma is heated to temperatures of 2\,MK as the eruption proceeds, and a radial speed of 580--650\,\kms\ is estimated from EIS and AIA data at 13:22~UT. This complements the observations obtained by SPICE of the same northern region of the filament, showing a filamentary structure as well as emission in multiple temperatures, and with speeds of several hundreds of \kms.

\subsection{Energetic electrons as seen in X-rays with STIX}
\label{sec-Sources}

\begin{figure*}
    \centering
    \includegraphics[width=0.95
    \textwidth]{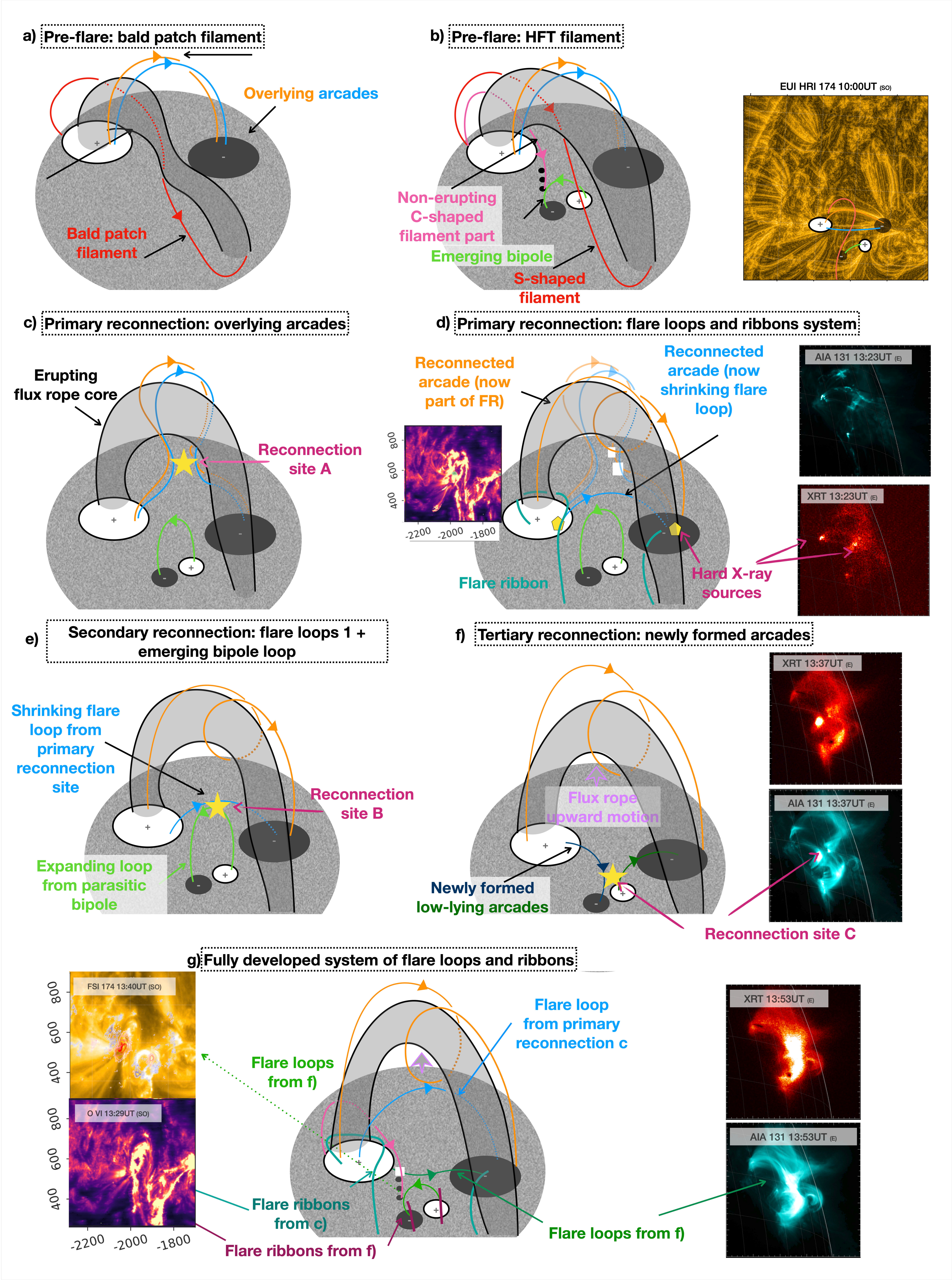}
    \caption{Summary diagram showing the three reconnection events involving the overlying arcades as well as the underlying arcades of the parasitic bipole. Some observations from SDO/AIA, \hinode\ XRT and \solo\ EUI/HRI and SPICE are shown as links to the features shown in the diagrams. Panel a) is related to the pre-flare magnetic field configuration of the bald patch filament, prior to the emergence of the parasitic polarity. Panel b shows the set of field lines describing the flux rope (red lines) and its core (grey cylinder), the overlying arcades (blue and orange), the emerging parasitic bipole (green) and the part of the northern filament that is seen after the eruption (pink). Panels b and c focus on the primary reconnection as expected from the Standard Flare Model in 3D, while panels d and e show the implication of reconnection involving the parasitic bipole. Panel f shows the complex flare ribbons and loops signatures observed in the dataset. 
    }
    \label{reconnection-diagram}
\end{figure*}

We employed the pixelated (information on individual detectors and pixels contained in the L1 `pixel' data) nature of the STIX science data to reconstruct  the HXR sources. 

Here we focus on the non-thermal HXR emission (i.e. energies above 25\,keV) that emanates from the chromospheric footpoints and is generated by non-thermal electrons. STIX images were generated using the maximum entropy method \citep[MEM\_GE,][]{Massa2020}, where visibilities were computed after integrating the counts over the energy range 25 to 50 keV, and over time intervals with durations between two and three minutes that cover the three major HXR peaks. The time bins are indicated in Fig.~\ref{goes_stix_lc} with the vertical dashed lines. The location of the sources was determined using the pointing information provided by the STIX aspect system \citep{Warmuth2020b}. 

We show the sources overplotted on the SPICE rasters and the EUI/FSI 174\,\AA\ images in Fig.~\ref{spice-intensities}b. During the first HXR peak (near 13:20~UT at \solo), a single footpoint source is seen close to the centre of the field of view. It coincides with the northern part of the eastern flare ribbon seen in EUV. This source remains visible (but is decaying) over the next 10\,min. During the second HXR peak (near 13:25~UT), a new footpoint source has become dominant that is associated with the western EUV ribbon, near $-1800''$ longitude. We note that since this HXR peak is located just in the middle between the EUI frames at 13:20:50 and 13:30:50 UT, we have plotted the source on both images. Since the eastern footpoint source is still present during the second HXR peak, we now observe the usual pair of non-thermal footpoints. Comparing the source locations with the PHI/HRT magnetogram (also plotted in Fig.~\ref{spice-intensities}b) reveals that the HXR footpoints are located in regions of opposite polarity, as is predicted by the magnetic reconnection scenario of solar flares. The third, fainter HXR peak (near 13:30~UT) is dominated by a new footpoint on the western ribbon that is located further southwards. 

In summary, the evolution of the non-thermal STIX sources suggests two distinct reconnection locations: one in the northern part of the AR (associated with the first two HXR peaks), and one further towards the south (third peak).

\section{Interpretation of observations with 3D MHD models}
\label{sec-Discussions}

A summary of the scenario of the flare is given in Fig.~\ref{reconnection-diagram}, with the following steps and panels: 

a) A few days and up to a few hours before the flare, the filament is seen as a low-lying structure, as indicated in Fig.~\ref{magnetograms} and discussed in Sect.~\ref{sec-Flux_emergence}. This low-lying filament is interpreted as a bald-patch filament. While the northern leg of the filament is anchored in the positive polarity of AR 12976, the southern leg is anchored in a region of low magnetic field (see Fig.~\ref{overview-activeregions}). 

b) In the pre-flare conditions, the core of the flux rope is represented by the grey shaded cylinder, with one set of footpoints anchored in the positive polarity, and the other anchored in the negative field polarity extending southwards outside of the FOVs of the \solo\ high resolution channels of the remote-sensing instruments. The single red line represents one of the flux rope field lines that encircles the grey core. We also distinguished the non-eruptive, C-shaped filament part in pink. This remaining filamentary part was pointed out in Sect.~\ref{sec-Chromospheric-IRIS} and was seen before and after the flare in the IRIS FOV. The emerging `parasitic' bipole field lines are represented with the green lines. At this stage, the parasitic bipole is believed to be a precursor of the early filament rise phase, as discussed in Sect.~\ref{sec-Precursor}. This emergence leads to the filament `detaching' from a bald patch configuration to an HFT configuration (see Sect.~\ref{sec-Flux_emergence}). Finally, a set of arcades overlying the filament are represented in orange and blue. These different sets of field lines are also indicated in one of the EUI/HRI images so as to guide the eye, and are accompanied by the online movie showing the PHI/HRT magnetograms overlaid on top of the EUI/HRI images (\url{https://owncloud.ias.u-psud.fr/index.php/s/8ycFaMFg4CBtZEi}).
    
c) At the flare trigger, the overlying arcades reconnect at site A. This is the archetypical scenario of the standard flare model in 3D, with more details shown in, for example, Fig.~5 in \citet{Aulanier2012}. The parasitic bipole is still represented, but does not intervene in this sequence.

{d)} This primary reconnection leads to the formation of two sets of field lines: a set of field lines that wrap around the core of the flux rope (orange), and a set of field lines corresponding to the flare loops {(light blue)}. This main reconnection is accompanied by hot emission sources as seen in the AIA 131\,\AA\ images and XRT (shown as inserts on the right), and a pair of non-thermal HXR footpoint sources detected by STIX, as discussed in Sect.~\ref{sec-Sources}.

{e)} As the eruption goes on and the flux rope moves upwards, two sets of magnetic field lines have converging motion. {Next,} 
    the flare loops (in blue) of the main reconnection site shrink down, against the emerging magnetic field. 
    These two sets of field lines then reconnect at site B (indicated in pink).  
    
{f)} {This second reconnection leads} to a new pair of field lines (in dark blue and dark green). This reconnection process is linked to the formation of hot coronal loops seen, for instance, in the \ion{Ne}{viii} line with SPICE (as discussed in Sect. \ref{sec-SPICE}) as well as to the appearance of an additional HXR footpoint source {(Fig.~\ref{spice-intensities})}. Next, the region between the two sets of new field lines becomes magnetic flux-depleted because of the time-dependent removal of the eruptive flux rope. This drives sink flows that induce a third reconnection \citep[more details on the vortex and sink flows created by eruptive flares are given in][]{Zuccarello2017}. 
        
{g)} This tertiary reconnection then reforms loops situated in the parasitic emerging bipole accompanied by ribbons at their footpoints (pink ribbons), as well as another overarching, low-lying set of field lines. The change in the magnetic connectivity related to the quadrupolar reconnection at the tertiary stage is best seen in the structures of the emitting loops in the XRT and AIA 131\,\AA\ images (insert). 
    
In summary, we interpret, in the framework of present 3D MHD models of flares and eruptions, the overall evolution of the plasma and magnetic configuration, 
the flare loops and ribbons as seen in the different observations, from EUI/FSI 174\AA, to SPICE, to SDO/AIA and \hinode/XRT  as summarised in Fig.~\ref{reconnection-diagram}.  It comprises the set of field lines (orange and blue) from the primary reconnection site A, and accompanying flare ribbons in cyan (panel c), as well as the field lines from the tertiary reconnection site C (light and dark green) and one set of ribbons in pink (panel f).     

Multiple reconnection processes in complex regions such as quadrupolar ones are not rare. An example is given in \citet{Gary2004} where the authors investigate an eruptive flare and subsequent events in a quadrupolar topology. In their case, the initiation of the magnetic flux rope eruption was believed to be triggered first by breakout. In that scenario, the fluxes of the central and overlying systems remain closed. Another event in the form of the release of a highly twisted filament was studied in \citet{Williams2005}. Then, an emerging bipole was suggested to play an important role in weakening the tether of a near-to-instability filament, bringing a small amount of magnetic stress to push the magnetic field structure to a critical state. This scenario is similar to the present one, although multi-points observations of the filament suggest that the 2 April eruption is not due to a highly kinked flux rope. A further analysis on the instability criteria (e.g. kink unstable vs. torus unstable) would be needed. Other examples of emerging flux triggering flux rope eruption can be even more complex than the situation described above \citep[e.g.][]{Louis2015} with forced reconnection between the expanding CME structure and neighbouring magnetic field {that} can also lead to long duration events \citep[e.g.][]{Goff2007}. 

{Another point to discuss is that the question of the destabilisation from the emerging bipole is not clearly evident. In the present scenario, we have considered that the emerging field is simply pushing the overlying flux rope up (stages a and b). However, reconnection between emerging magnetic structures and overlying fields also happens, as shown for example in numerical simulations such as those of \citet{Galsgaard2007} and \citet{Leake2014}. In the present case, this would ultimately lead to changes of connectivity in both the emerging flux and the pre-existing flux rope anchor points, and the amount of helicity carried by each structure. This, in turn, may have consequences in the observations found, such as the shapes of the ribbons, or the thread-like nature of the outflows, as discussed for instance in Fig. \ref{spice-doppler}. To better characterise the interaction between the emerging bipole and the overlying flux rope, one possibility is to use the information on the magnetic fields provided by both Hinode/SOT and PHI for a non-linear force-free field extrapolation, as an initial condition to a dynamical simulation of the magnetic fields during the eruption. This is however out of the scope of the paper, and is left aside for further investigation.}

Atypical flares such as the one studied in \citet{Dalmasse2015} showed the importance of considering the 3D magnetic volume and the topological signatures to decipher the mechanisms at the heart of flares that do not fit within the typical, standard flare model. In the present case, the combination of multiple points of view and diagnostics made understanding the role of the complex quadrupolar topology possible.

\section{Conclusions}
\label{sec-Conclusions}

The 2 April 2022 M-class eruptive flare was the first eruptive flare caught simultaneously by all remote-sensing instruments onboard the \solo\ spacecraft (except Metis), and with several Earth-orbiting assets (the heliospheric imager on board the mission, SoloHI, did also monitor a CME which is the subject of a subsequent study). Because of the location of the spacecraft in relation with Earth (shown in Fig. \ref{solo-position}), this event provided for the first time ever the possibility to obtain an extensive diagnostic of the flaring region via the help of three spectrometers positioned at different longitudes, complemented by measurements of the magnetic field and hard X-rays. 

Thanks to the long term coverage of the region by the \solo\ remote-sensing instruments (see the time coverage in Fig. \ref{xrayflux}), we were able to catch the pre-eruptive evolution of the region. Both SDO/HMI and \solo/PHI/HRT were crucial in detecting the emergence of a parasitic polarity (Fig. \ref{magnetograms}) that plays an important role in triggering the filament eruption. 

We observed several pre-eruptive activities: from enhanced non-thermal velocity in the filament region (Fig. \ref{vnt-eis}), to the apparent motion of overlying arcades above the filament during its slow rise (Figs. \ref{hri-spice-loops-stackplots} and \ref{aia-looptop}). This provides several pieces of evidence that  activation processes can be observed a few hours before a flaring event and can be used as a warning before the flare takes place. It is compatible with models that put forwards a filament-bearing {magnetic configuration} being brought up to an unstable state via photospheric/lower corona activity. 

IRIS at Earth's orbit showed that not all the filament actually erupts during the flaring event (see Fig. \ref{iris}), with a portion of the northern part remaining anchored at the negative leading polarity of AR 12976. SPICE on \solo\ provided intensity and Doppler maps every 15~min.  Figures \ref{spice-intensities} and \ref{spice-fexviiimaps} showed a complex network of flare loops and ribbons, related to the reconnection processes taking place in the quadrupolar configuration due to the presence of the parasitic bipole. Figure \ref{spice-doppler} showed the blueshifted components of the filament's northern leg, with velocities reaching up to 400~\kms. EIS on \hinode\ {showed} blue-shifted emission in the transition region and coronal lines in the northern leg of the flux rope prior to the flare peak, with radial velocities of $\approx 300$ to $\approx 800$~\kms. The motion detected in the AIA coaligned images, along with the Doppler-shifted \ion{He}{ii} emission are suggestive of the unwinding motion of the flux rope as it expands (Fig. \ref{fig.movie1}). Finally, the development of hot post-flare loops is demonstrated with the emission intensities of \ion{Fe}{xxiv}. 

The co-aligned STIX data and intensity maps from both SPICE and EUI/FSI (Fig. \ref{spice-intensities}b) show that the non-thermal emissions are co-spatial with specific locations along the flare ribbons.

These different sets of observations provide for the first time a multipoint spectroscopic measurement of a flaring region, not only allowing multi-thermal diagnostics of the different parts of the region (from the ribbons to the flare loops and the filament material), but also constraining the velocity of the erupting structure.

The  magnetograms provided by PHI/HRT are invaluable in complementing the data from HMI. In particular, due to the late emergence of the parasitic bipole, the evolution captured by PHI/HRT was almost missed by HMI, and not well constrained due to the position on the western limb (see Fig. \ref{magnetograms}). The overlays of the magnetograms with the FSI images obtained of the flare loops shown in Fig. \ref{spice-intensities}b were crucial to understand the anchor of these loops, and therefore understand the different steps leading to the apparition of the different structures. The addition of the HXR sources seen by STIX offered sharp observations of non-thermal emissions, and further insight on the connectivity of the loops especially when overlaid with the flare ribbons (Fig. \ref{spice-intensities}a).

This extensive set of observations has allowed us to {study and propose an explanation for} the complex multi-step reconnection process taking place during this M-class flare, as {shown} in Fig. \ref{reconnection-diagram}. Although seemingly typical of usual flares, the current event, with its complex network of ribbons and loops, showed the importance of taking into account the parasitic bipole in both triggering the filament eruption, and reconnecting with the erupting filament, leading to the formation of secondary and tertiary flare loops and ribbons systems. Understanding these complex observations with multiple reconnection sites within the context of 3D MHD models was only possible due to the complementary nature of the data. 

This first extensively observed flare with the \solo\ mission also provides some insights on what is yet to come in the following years. While this observation was made in the first nominal mission perihelion campaign, similar active region follow-up and `flare watch' observing plans will continue to be proposed for future campaigns. With the added coordination with external observatories, both in space and on ground, more insights will be gained by such multi-point, multi-instrument observations. 

\begin{acknowledgements}
    Solar Orbiter is a mission of international cooperation between ESA and NASA, operated by ESA. We thank the \solo\ instrument teams for making the data accessible, and for \hinode, IRIS for coordinating the observations during the `long-term AR' SOOP. 
    The development of the SPICE instrument was funded by ESA and ESA member states (France, Germany, Norway, Switzerland, United Kingdom). The SPICE hardware consortium was led by Science and Technology Facilities Council (STFC) RAL Space and included Institut d’Astrophysique Spatiale (IAS), Max-Planck-Institut für Sonnensystemforschung (MPS), Physikalisch-Meteorologisches Observatorium Davos and World Radiation Center (PMOD/WRC), Institute of Theoretical Astrophysics (University of Oslo), NASA Goddard Space Flight Center (GSFC) and Southwest Research Institute (SwRI). The in-flight commissioning of SPICE was led by the instrument team at UKRI/STFC RAL Space. A.F.’s and A.G.’s research is funded by UKRI STFC.
    The EUI instrument was built by CSL, IAS, MPS, MSSL/UCL,
PMOD/WRC, ROB, LCF/IO with funding from the Belgian Federal Science Policy
Office (BELSPO/PRODEX PEA 4000134088, 4000112292, 4000117262, and 4000134474). 
    The STIX instrument is an international collaboration between Switzerland, Poland, France, Czech Republic, Germany, Austria, Ireland, and Italy.
    The German contribution to SO/PHI is funded by the BMWi through DLR and by MPG central funds. The Spanish contribution to SO/PHI is funded by FEDER/AEI/MCIU (RTI2018-096886-C5) and a “Center of Excellence Severo Ochoa” award SEV-2017-0709 to IAA-CSIC. The French contribution is funded by CNES.
    LBR acknowledges financial support from the Spanish MCIN/AEI 10.13039/501100011033 through projects RTI2018-096886-B-C5, PID2021-125325OB-C5, and the "Center of Excellence Severo Ochoa" award CEX2021-001131-S to Instituto de Astrof\'{\i}sica de Andaluc\'{\i}a, all co-funded by European Regional Development Funds.  
    \hinode\ is a Japanese mission developed and launched by ISAS/JAXA, with NAOJ as domestic partner and NASA and STFC (UK) as international partners. It is operated by these agencies in cooperation with ESA and NSC (Norway). IRIS is a NASA small explorer mission developed and operated by LMSAL with mission operations executed at NASA Ames Research Center and major contributions to downlink communications funded by ESA and the Norwegian Space Centre.

    This work was supported by CNES (via APR -BC4500072977), focused on SPICE and EUI on Solar Orbiter. We also acknowledge MEDOC (Multi Experiment Data \& Operation Center), a National Center for Space Solar Physics Data, approved by CNES, in the frame of an agreement between CNRS / INSU, Université Paris-Saclay and CNES. MEDOC is located at Institut d'Astrophysique Spatiale in Orsay. 
    This work was supported by a Paris Ile-de-France Region DIM ORIGINES contract (22HR2755, S.M.).
    P.R.Y. acknowledges support from the GSFC Individual Scientist Funding Model competitive work package and the \hinode\ project. D.M.L. is grateful to the Science Technology and Facilities Council for the award of an Ernest Rutherford Fellowship (ST/R003246/1). F.S. acknowledges support by the German Space Agency (DLR) grant No. 50OT1904. D.B. is funded under STFC consolidated grant number ST/S000240/. The work of D.H.B. was performed under contract to the Naval Research Laboratory and was funded by the NASA \hinode\ programme.  A.N.Z. thanks the Belgian Federal Science Policy Office (BELSPO) for the provision of financial support in the framework of the PRODEX Programme of the European Space Agency (ESA) under contract number 4000136424.
    We recognise the collaborative and open nature of knowledge creation and dissemination, under the control of the academic community as expressed by Camille No\^{u}s at http://www.cogitamus.fr/manifesteen.html. This research made use of several open-source software packages including SunPy \citep{sunpy_community2020},  sunraster, AstroPy \citep{astropy:2022}, and matplotlib \citep{Hunter:2007}.
\end{acknowledgements}

\appendix
\section{Complementary EIS and AIA data analysis explanation}

\subsection{EIS and AIA alignment}
\label{eis_aia_alignement}

   To compare the EIS data with AIA 304\,\AA, 
the two datasets were spatially aligned by extracting slits of emission from the AIA image sequence in order to reproduce the time-$y$ image from the EIS \ion{He}{ii} 256.32\,\AA\ line. The IDL routine \textsf{aia\_make\_eis\_raster} was used for the alignment and is available from the GitHub repository \href{https://github.com/pryoung/sdo-idl}{pryoung/sdo-idl}. Further details are available in EIS Software Note No.~26 \citep{young_peter_r_2023_7582257}. The best alignment was found by adding an offset of $(+8$\as,$+15$\as) to the EIS data in addition to the standard EIS--AIA offset obtained from the IDL routine \textsf{eis\_aia\_offsets}. All EIS pointings discussed in Sect.~\ref{sec-eisdoppler} include this correction.

\subsection{LOS to radial speeds conversion}
\label{los_radial_speeds}

Doppler shifts of the EIS emission lines yield line-of-sight (LOS) plasma speeds, while the tracking of intensity features in the AIA images yields plane-of-sky (POS) speeds. Combined, the two speeds can then give the true plasma velocity, but given the complexity of the filament evolution it can be difficult to associate an imaged feature with a spectral feature. An alternative is to assume the filament erupts radially and then the POS or LOS speed can be converted to the plasma velocity. The position of the filament projected back onto the solar disk can be made based on the flare ribbons that are assumed to be underneath the filament {and near the photospheric level}. Comparing an AIA 171\,\AA\ image from 13:35~UT with the SPICE images at this time (Fig.~\ref{spice-intensities}), we estimate the position angle to be 65$^\circ$ (from the Earth-Sun line). Thus the LOS and POS speeds should be multiplied by 2.4 and 1.1, respectively, to estimate the radial speeds.

\section{SPICE astigmatism and its effect on the Doppler analysis}
\label{spice-astigmatism}

The astigmatism of the SPICE optics produces false velocity signals that are correlated with the intensity gradients in the image.  The primary source of these false velocity signals is due to intensity gradients along the slit, which is negatively correlated with apparent velocity.  There is also a smaller (positive) correlation with the intensity perpendicular to the slit.

\begin{figure}
\centering
\includegraphics[width= 0.49\textwidth]{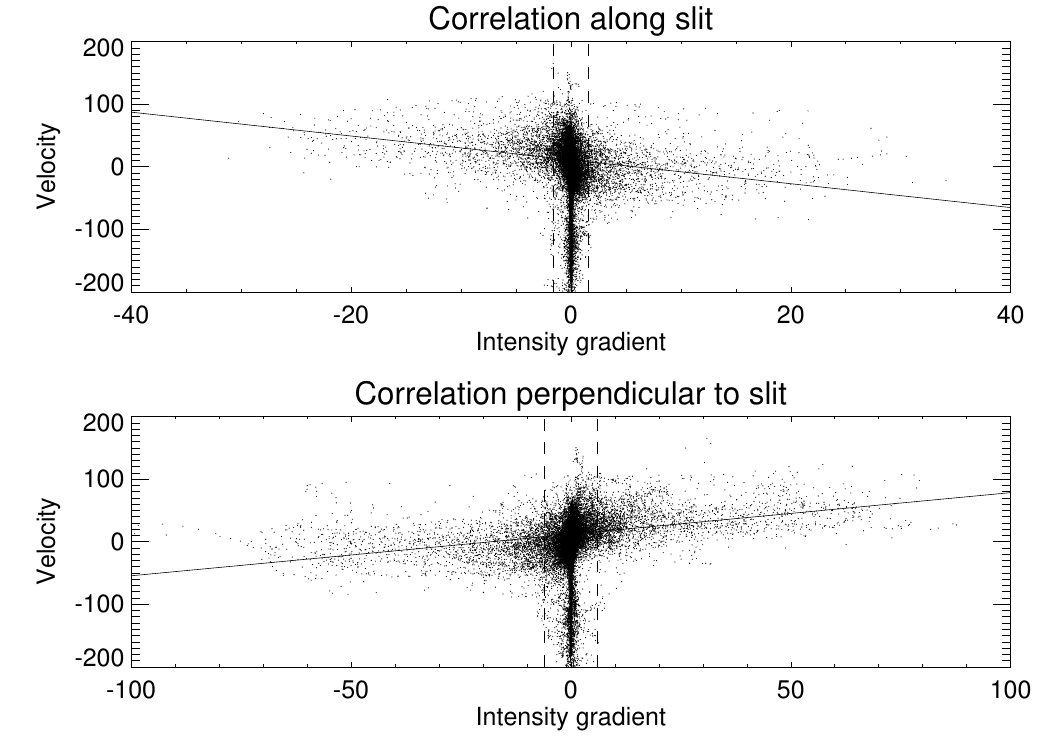}
\caption{Sample correlations between the intensity gradients along (top) or across (bottom) the SPICE slit with the velocity derived from line fitting.  The verticle dashed lines represent the $\pm 1 \sigma$ range that was excluded from the linear fit (solid line).  This sample was taken from the C~{\sc iii} raster taken between 13:15:36 and 13:20:57 UT.  The strong blueshift from the erupting filament is clearly visible between the dashed lines.} 
\label{spice_correlation}
\end{figure}

\begin{figure*}
\centering
\includegraphics[width= \textwidth]{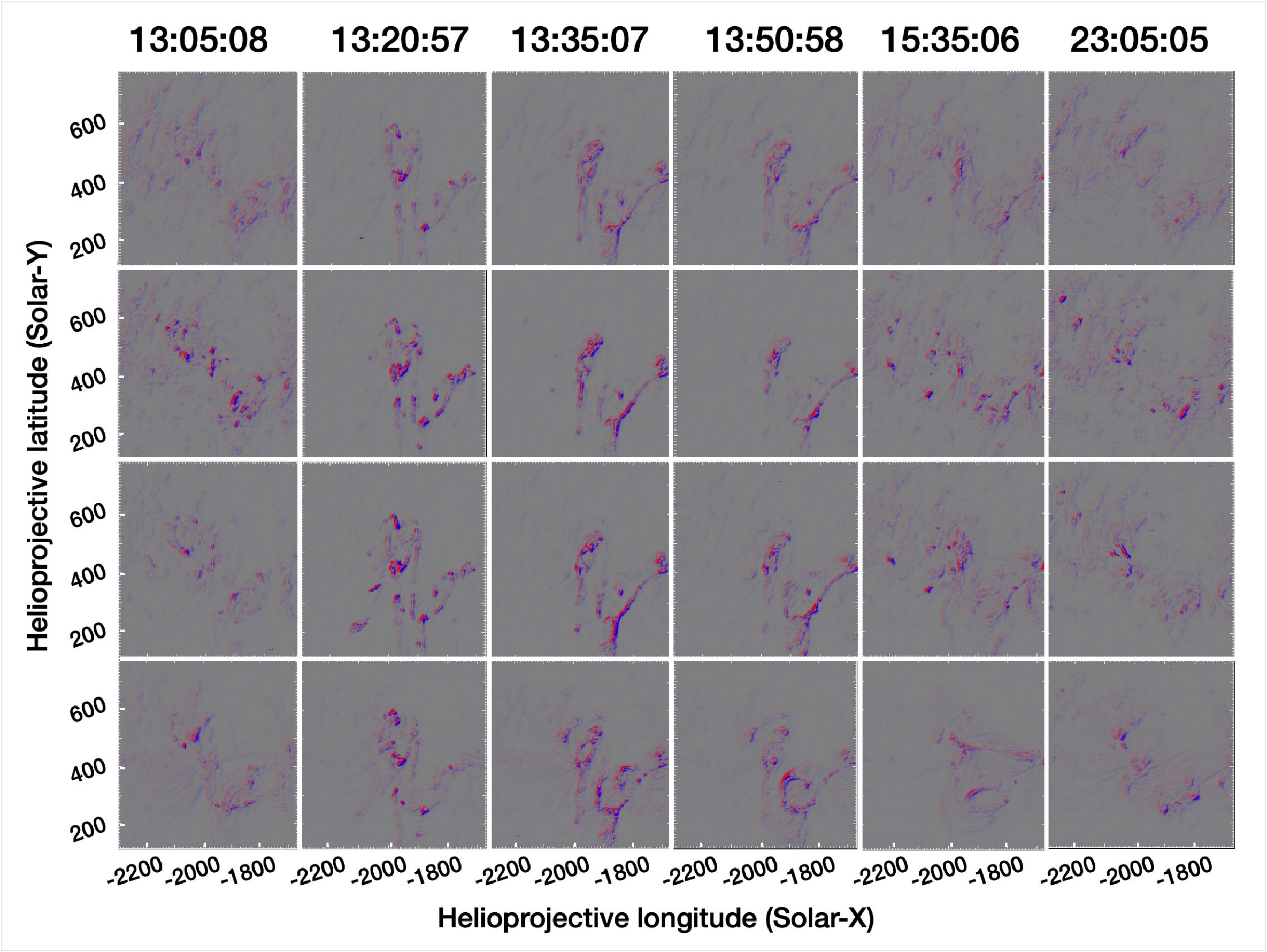}
\caption{Simulated Doppler velocity maps representing a proxy for the effect of the astigmatism in the SPICE optics on the velocities derived from line profiles in Figure~\ref{spice-doppler}. The times correspond to the observation time.}
\label{spice_proxy}
\end{figure*}

A simple IDL routine was developed to explore how the intensity gradients could be affecting Fig.~\ref{spice-doppler}.  The intensity gradient at a pixel location $j$ along the slit was estimated by subtracting the intensity at pixel $j-1$ from that at pixel $j+1$, and compared with the fitted velocity at pixel $j$.  Similarly, the intensity gradient at raster position $i$ was estimated by subtracting the intensity at position $i-1$ from that at $i+1$.  A sample comparison is shown in Fig.~\ref{spice_correlation}.  A linear fit was made along each dimension, excluding the region within 1$\sigma$ of zero gradient, where the real solar signal is assumed to lie.  We used the SSW routine \verb+poly_fit_most.pro+ to filter out extreme values, but the results do not differ much from what a simple linear fit would give for most images.  The slopes of these fits were then multiplied into the gradient along each dimension, and combined to form a proxy for the instrumentally generated velocity.  These proxy maps are shown in Fig.~\ref{spice_proxy}.

Examining Fig.~\ref{spice_proxy}, it is clear that the strong blue-shift of the erupting filament in Fig.~\ref{spice-doppler} has no counterpart in the instrumental proxy, and therefore can be safely taken as solar in origin.  The proxy images do show velocity signals at the locations of the flare ribbons that are similar to those in Fig.~\ref{spice_proxy}, and thus most are assumed to be instrumental in origin.  A possible exception is the strong redshift feature associated with the eastern primary ribbon in the 13:15:36--13:20:57 data; the simulation shows both red and blue shifts, but the data show only redshift, and much stronger than the simulation.  Similarly, the strong redshift features seen in the 15:29:46--15:35:06 data do not seem to have an obvious counterpart in the simulated data.  However, no attempt is made to interpret these features in the present work.

\bibliographystyle{aa} 
\bibliography{biblio} 
\end{document}